\pdfoutput=1
\documentclass[aps,twocolumn,a4paper,showpacs,superscriptaddress,
    floatfix]{revtex4-1}
\usepackage{graphicx}
\usepackage{siunitx}
\usepackage{amsmath,amssymb}
\usepackage{hyperref}
\usepackage{datetime}
\usepackage{enumitem}

\usepackage[dvipsnames]{xcolor}
\usepackage{xfrac} %sfrac
\usepackage{tikz}

\hypersetup{
    colorlinks,
    linkcolor={red!50!black},
    citecolor={blue!50!black},
    urlcolor={blue!80!black}
}

\newcommand{\e}[0]{\ensuremath{\epsilon}}
\newcommand{\q}[1]{\ensuremath{\langle #1 \rangle}}
\newcommand{\pa}[1]{\left(#1\right)}
\newcommand{\bra}[1]{\left[#1\right]}
\newcommand{\pfrac}[2]{\pa{\frac{#1}{#2}}}

\begin{document}

\title{Band and correlated insulators of cold fermions in a mesoscopic lattice}

\author{Martin Lebrat}
\affiliation{Department of Physics, ETH Zurich, 8093 Z\"{u}rich, Switzerland}
\author{Pjotrs Gri\v{s}ins}
\affiliation{Department of Quantum Matter Physics, University of Geneva, 1211 Gen\`{e}ve, Switzerland}
\author{Dominik Husmann}
\author{Samuel H\"{a}usler}
\author{Laura Corman}
\affiliation{Department of Physics, ETH Zurich, 8093 Z\"{u}rich, Switzerland}
\author{Thierry Giamarchi}
\affiliation{Department of Quantum Matter Physics, University of Geneva, 1211 Gen\`{e}ve, Switzerland}
\author{Jean-Philippe Brantut}
\affiliation{Institute of Physics, \'{E}cole Polytechnique F\'{e}d\'{e}rale de Lausanne, 1015 Lausanne, Switzerland}
\author{Tilman Esslinger}
\affiliation{Department of Physics, ETH Zurich, 8093 Z\"{u}rich, Switzerland}

\date{\pdfdate}

\begin{abstract}
We investigate the transport properties of neutral, fermionic atoms passing through a one-dimensional quantum wire containing a mesoscopic lattice.
The lattice is realized by projecting individually controlled, thin optical barriers on top of a ballistic conductor.
Building an increasingly longer lattice, one site after another, we observe and characterize the emergence of a band insulating phase,
demonstrating control over quantum-coherent transport.
We explore the influence of atom-atom interactions and show that the insulating state persists
as contact interactions are tuned from moderately to strongly attractive.
Using bosonization and classical Monte-Carlo simulations we analyze such a model of interacting fermions and find good qualitative agreement with the data. The robustness of the insulating state supports the existence of a Luther-Emery liquid in the one-dimensional wire.
Our work realizes a tunable, site-controlled lattice Fermi gas strongly coupled to reservoirs, which is an ideal test bed for non-equilibrium many-body physics.
\end{abstract}

\pacs{
}

\maketitle

\begin{figure}[t!]
    \includegraphics{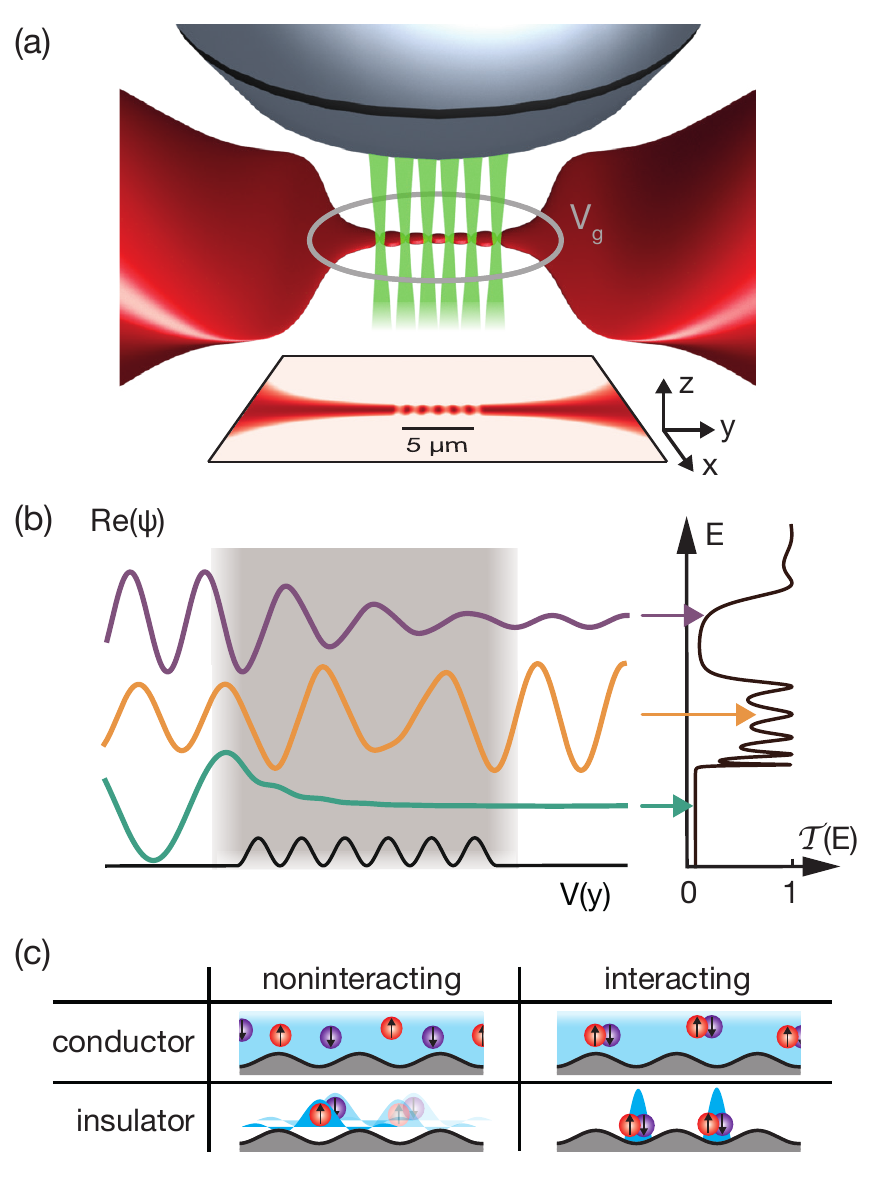}
    \caption{Concept and experimental realization. (a) Sketch of a one-dimensional lattice projected onto a quantum wire that connects two macroscopic atom reservoirs. The lattice beam is here made up of six repulsive, holographically-shaped barriers that are imaged into the atomic plane with a high-resolution microscope. The sites are spaced by $\SI{0.97}{\micro\meter}$. An attractive gate beam allows us to locally increase the chemical potential by an energy $V_g$.
    (b) Real part of the wavefunction $\psi(y)$ and transmission $\mathcal{T}$ for a single atom incoming on a one-dimensional lattice of barriers for various energies $E$.  $\mathcal{T}(E)$ is zero for energies below the lattice zero-point energy (green arrow), and features a conduction band (yellow) and a band gap (purple). The non-zero transmission in the gap as well as the modulation in the conduction band originate from the finite size.
    (c) Possible insulating and conducting states for two-component fermions in a one-dimensional lattice. The atoms are delocalized at incommensurate densities and become localized at lattice fillings close to two particles per site. For strong attractive interactions, the conductor is made of extended pairs, and the insulator is made of pairs pinned to lattice sites. For non-interacting particles, the system forms a band conductor and a band insulator for incommensurate and commensurate fillings, respectively.
    }
    \label{fig:setup}
\end{figure}

\begin{figure*}
    \includegraphics{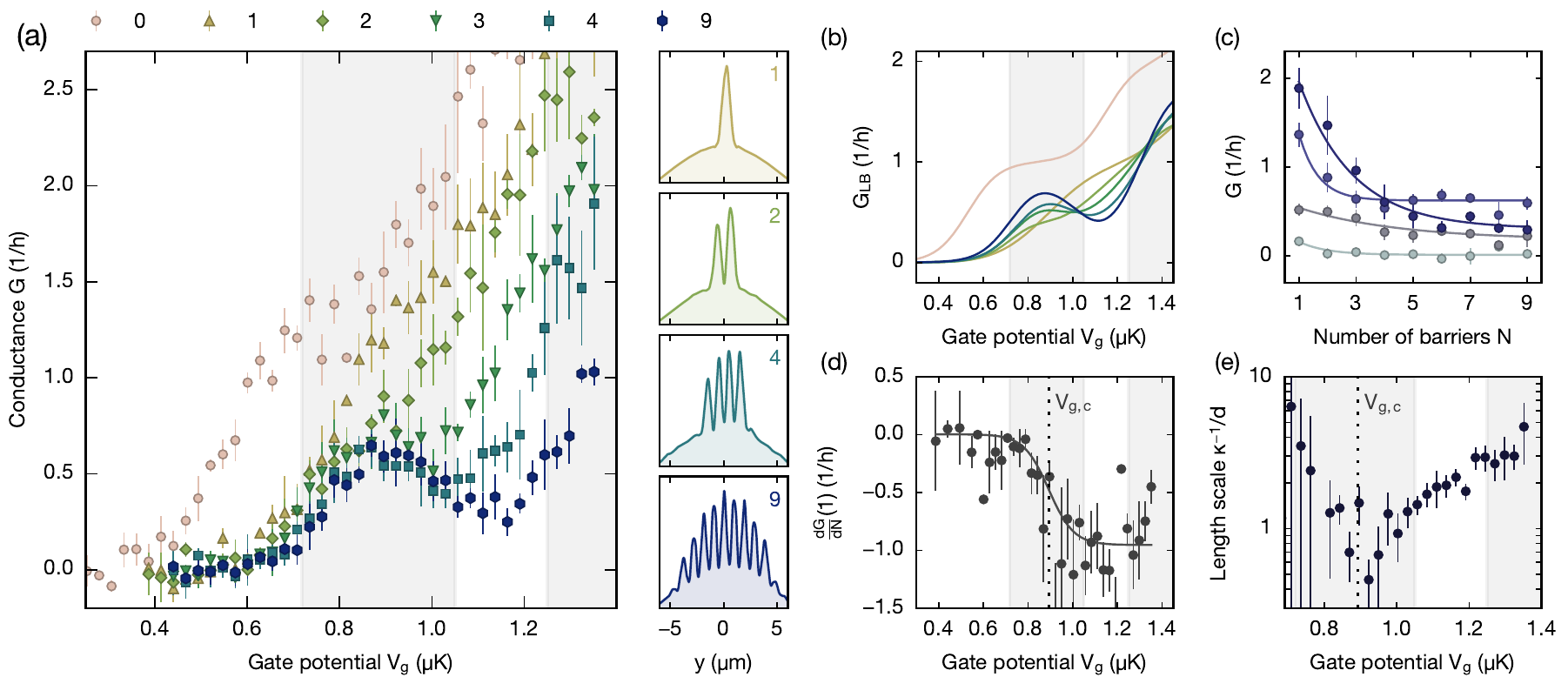}
    \caption{Building up a lattice site by site. (a) Conductance $G$ as a function of attractive gate potential $V_g$ without scattering structure; with one barrier; with a lattice of 2 to 9 barriers. The potential height of the central barrier is set to $V_l = \SI{0.40(2)}{\micro\kelvin} = 0.94(5) E_r$, where $E_r$ is the lattice recoil energy. The first and second bands of an infinite lattice of same height are indicated by gray areas, separated by a gap which coincides with the location of the conductance minimum for 9 barriers. Right panels: actual quasi-one-dimensional potentials for 1, 2, 4 and 9 barriers. (b) Conductance $G_\text{LB}$ obtained from non-interacting Landauer-B\"{u}ttiker theory at a temperature $T = \SI{67}{\nano\kelvin}$ and a typical chemical potential difference $\Delta \mu = \SI{0.1}{\micro\kelvin}$ between the reservoirs. (c) Conductance $G$ as a function of number of barriers $N$ for four gate potentials $V_g = \SI{0.55}{\micro\kelvin}, \SI{0.73}{\micro\kelvin}, \SI{0.95}{\micro\kelvin}$ and $\SI{1.11}{\micro\kelvin}$ (light to dark blue), together with exponential fits $G(N) = (G_1 - G_{\infty}) \exp[-\kappa (N-1)] + G_{\infty}$. (d) Fitted derivative of the conductance for one lattice site $G'(1)$. A sharp drop is located at $V_{g,c} = \SI{0.85(2)}{\micro\kelvin}$ using a sigmoid fit, hinting that transport is becoming non-ballistic. (e) Characteristic length scale $\kappa^{-1}$ associated with the exponential decay and normalized by the lattice spacing $d$. It reflects the inverse imaginary part of the wavevector of a Bloch wave in the lattice gap (white area).}
    \label{fig:site_number}
\end{figure*}

\section{Introduction}

Quantum effects are the cornerstone of most of the properties encountered in electronic materials and devices. Even the fundamental and apparently simple question as to why a material behaves in a transport experiment as a metal, an insulator or a semiconductor, relies on subtle effects, such as the antisymmetrization of fermionic wavefunctions and the interference of matter waves in a periodic potential \cite{ziman_1972}. While non-interacting systems are mostly well understood, predicting the transport properties when interactions are relevant often remains an intellectual challenge, which is of practical importance, as it paves the way to engineering devices with new functionalities.

Given the complexity of interacting many-body problems, simplifying theories have been worked out, which are particularly suitable for high dimensional systems. For example, fermions with weak repulsive interactions can be described by Landau's Fermi liquid theory, where the many-body excitations simply behave like free particles, the so-called Landau quasi-particles, whose effective parameters, like the mass, are renormalized. For weak attractive interactions, such a Fermi liquid is unstable due to the formation of pairs, which ultimately turns it into a superfluid that can be described by mean-field theory. However, there are situations where such simplifying assumptions are not applicable. This is particularly the case in one-dimensional interacting quantum systems, where Fermi liquid or mean-field descriptions fail and where all excitations become collective, leading to a set of properties known as a Tomonaga-Luttinger liquid \cite{giamarchi_quantum_2003}.

One of the important challenges in the context of transport is the robustness of the superfluid state against scattering events and perturbations happening at the single particle level. In two or three dimensions, one expects for very strong scattering a superfluid-to-insulator transition to take place. This may for instance arise from the competition between superfluidity and the presence of a periodic lattice causing Bragg scattering at the band edges. Indeed, calculations based on mean-field theory support the occurrence of a superfluid to band-insulator transition as the strength of the periodic lattice is increased \cite{nozieres_semiconductors_1999}.

In one dimension, however, a very different outcome is expected. Based on work by A. Luther and V.J. Emery \cite{luther_backward_1974}, it is known that a one-dimensional fermionic system with attractive contact interactions \emph{always} exhibits an excitation gap in the spin sector. The charge sector remains gapless and can be described as a Tomonaga-Luttinger liquid of spinless pairs, which can be viewed as composite bosons similar to Cooper pairs. Therefore, the presence of a weak periodic lattice and the correct commensurability, i.e. one pair per site, leads to the opening of a gap in the charge sector. This turns the system into a correlated insulator, even for arbitrarily strong attractive interactions. This is in stark contrast with the mean-field predictions applicable for higher dimensions.

Studying the effect of a weak periodic structure on the transport properties of fermions with an attractive contact interaction would thus address this challenge and at the same time serve as a probe for the existence of the Luther-Emery liquid and its properties. This is the task that we undertake in the present paper, using and expanding the toolbox of cold-atom experiments \cite{kaufman_two-particle_2014,murmann_two_2015,barredo_atom-by-atom_2016,endres_atom-by-atom_2016} to investigate transport in mesoscopic lattices \cite{seaman_atomtronics_2007,bruderer_mesoscopic_2012}. Thanks to our ability to optically imprint an arbitrary structure on a one-dimensional wire between two large atom reservoirs [see Fig.~\ref{fig:setup}(a)], we perform conductance measurements through a weak periodic potential as a function of the chemical potential. With reservoirs in the normal phase, we observe the emergence of a band structure as the number of individually controlled and equidistantly positioned scatterers is increased [Fig.~\ref{fig:setup}(b)]. In addition to changing the density in the wire, the conductor-to-insulator transition is further characterized by tuning the lattice height and temperature. We then increase the attractive interactions to unitarity and still observe an insulating phase at commensurate filling, with a conductance very close to the one observed in the normal system, indicating a crossover from a band insulator to a correlated insulator. The persistence of the insulating behavior even for resonant interactions and superfluid reservoirs is a strong indication of the existence of a Luther-Emery liquid pinned on the weak periodic potential [Fig.~\ref{fig:setup}(c)]. From a more general perspective, our work extends recently developed methods for conductance measurements with cold atoms \cite{krinner_two-terminal_2017,husmann_connecting_2015,valtolina_josephson_2015,krinner_mapping_2016,hausler_scanning_2017} to unexplored regimes of strongly correlated insulators.

The plan of the paper is as follows. After detailing the experimental setup in Section~\ref{sec:setup}, we present measurements of the conductance as a function of the chemical potential in the wire and the number of sites in the lattice in Section~\ref{sec:interf}. We then study the conductance of a fixed-length lattice as a function of lattice height and temperature in Section~\ref{sec:condins}, showing the boundaries between band conductor and band insulator. In Section~\ref{sec:interactions} we measure the conductance as a function of chemical potential for various interaction strengths and compare the outcome with the results of a Tomonaga-Luttinger liquid model. Finally, we present our conclusions in Section~\ref{sec:conclusions}. Technical details can be found in the Appendix.

\section{Setup} \label{sec:setup}

The structure central to our experimental setup is a quantum wire smoothly connected to two reservoirs acting as a source and a drain \cite{krinner_observation_2015}, typically containing altogether $N=9 \cdot 10^4$ $^6$Li atoms in each of the lowest and third lowest hyperfine states [Fig.~\ref{fig:setup}(a)].
The wire is created by intersecting the dark planes of two orthogonal, repulsive, $\text{TEM}_{01}$-like laser beams. The vertically (resp. horizontally) propagating beam has a Gaussian envelope with a $1/e^2$ waist of $\SI{9.1(3)}{\micro\metre}$ (resp. $30(1)\,\SI{}{\micro\metre}$).
They confine atoms to a quasi-one-dimensional geometry with transverse frequencies $\omega_x = 2\pi \cdot 22(9) \, \SI{}{\kilo\hertz}$ and $\omega_z = 2\pi \cdot 13(5)\,\SI{}{\kilo\hertz}$ respectively at the center.
We tune the local chemical potential in and around the wire using an attractive Gaussian `gate' beam, propagating along the $z$ direction with a $42.5(3) \, \SI{}{\micro\metre}$ waist and creating a potential dip $V_g$ at its center.
In our measurements, this varies the local density from zero to about 3 atoms per micron, which amounts to populate up to two transverse modes of the wire.
The reservoirs are evaporatively cooled down to an absolute temperature $T = 67(3) \SI{}{\nano\kelvin}$, which is one order of magnitude smaller than $\hbar \omega_x/k_B$ and $\hbar \omega_z/k_B$. The cooling is efficiently performed in a homogeneous magnetic field $B = \SI{690}{G}$, at which the three-dimensional scattering length $a$ diverges due to a broad Feshbach resonance. This choice compels us to subsequently ramp up the magnetic field to the right side of the resonance, where the scattering length is tuned to its transport value, from $a = -\infty$ to $a=-2.65 \cdot 10^3 \,a_0$, where $a_0 =  5.29\cdot10^{-11}$~m is the Bohr radius.

Finally, the mesoscopic lattice is produced using light at $\SI{532}{\nano\meter}$  holographically shaped with a Digital Micromirror Device (Appendix~\ref{A:Experimental details}). We can imprint up to nine barriers centered on the wire and equally spaced by $d = \SI{0.97}{\micro\meter}$ [Fig.~\ref{fig:setup}(a)], associated to a lattice recoil energy $E_r = \hbar^2 (\pi/d)^2 /2 m = \SI{0.42}{\micro\kelvin}$, where $m$ is the mass of $^6\text{Li}$ atoms, setting here and in the following $k_B = 1$.
The setup gives full control over the number and positions of the barriers, allowing us to change the potential landscape at the single site level.
This is similar in spirit to techniques pioneered with scanning tunneling microscopes on solid-state surfaces \cite{eigler_positioning_1990,khajetoorians_atom-by-atom_2012}, yet going beyond non-interacting situations.

\section{Interferences and band gap opening} \label{sec:interf}

A fundamental consequence of quantum mechanics on transport is the role played by interferences in determining the conductance of a system.
For periodically arranged barriers, constructive or destructive interferences in the forward direction yields the well-known lattice band structure even when scattering by an individual barrier is very weak, which results in the energy-dependence of the transmission of an incident matter-wave \cite{cheiney_realization_2013}.
The transmission probability, which is essentially a single-particle property, is in turn proportional to the linear response conductance of a non-interacting Fermi gas through the Landauer-B\"{u}ttiker formula. 

We observe how transport is hampered by a lattice built site by site, by measuring the conductance $G$ of the system for increasing number of barriers as a function of the local chemical potential in the wire $\mu_\text{wire} = \mu_\text{res} + V_g - V_\text{wire}$, where the reservoir chemical potential $\mu_\text{res}$ and the potential of the wire $V_\text{wire}$ stay typically constant. Varying the gate potential $V_g$ therefore allows us to probe the conductance in an energy-resolved manner.

These measurements are performed with an attractive Fermi gas with a scattering length of $a=-2.65 \cdot 10^3 \,a_0$, where we expect the gas to be normal in the reservoirs and the wire. The reservoirs are characterized in the bulk by the dimensionless parameter $1/k_F a = -2.1$, where $k_F = \sqrt{2 m \mu_\text{res}}/\hbar$ is the Fermi wavevector and $\mu_\text{res} = \SI{0.47}{\micro\kelvin}$ the typical chemical potential.
This parameter locally increases in the three-dimensional regions close to the wire due to the presence of the attractive gate beam, but temperature remains above the local superfluid critical temperature \cite{haussmann_thermodynamics_2007} for the values of the gate potential $V_g < \SI{1.5}{\micro\kelvin}$ reached in our experiment.
At the center of the wire, the spin gap $\Delta_s$ is also smaller than the temperature. One-dimensional theory based on Bethe ansatz in the absence of any scattering potential \cite{Fuchs2004} predicts values $\Delta_s < \SI{30}{\nano\kelvin}$ for densities larger than 1 atom per micron (Appendix \ref{A:Data analysis}).

The results are shown in Fig.~\ref{fig:site_number}(a).
As long as the gate potential is not large enough for the local chemical potential $\mu_\text{wire}$ to be positive, conductance is zero.
In the barrier-free wire, $G$ increases with $V_g$, as an increasing number of transverse modes gets populated.
% A shoulder appears around $V_g = \SI{0.8}{\micro\kelvin}$ as a remnant of the first conductance plateau of a weakly interacting Fermi gas \cite{krinner_observation_2015}, at a conductance $G \approx 1.5/h$ higher than the conductance quantum $1/h$, an effect of weak attraction between the particles \cite{krinner_mapping_2016}.
For a Fermi gas with the same attractive scattering length flowing through a quantum point contact, a plateau would be expected at a conductance higher than the conductance quantum $1/h$ as an effect of attraction between the particles \cite{krinner_mapping_2016}.
In contrast, it does not appear here, most likely owing to the choice of a longer wire, which was shown to produce less robust conductance plateaus \cite{krinner_observation_2015}.

Upon placing a single repulsive barrier of height $V_l = \SI{0.40(2)}{\micro\kelvin}$ at the center of the wire, the conductance curve is shifted to the right, indicating that larger gate potentials $V_g$ are required to overcome the barrier potential $V_\text{wire}$ and enable transport. For increasing gate potential, $G$ increases monotonically up to high values characteristic of the multimode regime. As more barriers are inserted to form a lattice with a fixed central height $V_l$, the conductance at low gate potential is barely modified up to $V_g \approx \SI{0.9}{\micro\kelvin}$. This is in contrast with classical Ohm's law for series addition of incoherent barriers, where conductance is expected to decrease as the inverse of the length.

Beyond this value of $V_g$, transport is significantly modified and an inflection point appears already with three barriers, which then turns into a local conductance minimum for larger numbers of barriers. The reduction of conductance with increasing chemical potential is characteristic of hole-like transport, and signals the emergence of the band-insulating state. This effect is observed here for a shallow lattice potential, $V_l = 0.94(5) E_r$, where the band gap and bandwidth of the infinite, homogeneous lattice with equivalent spacing and height are expected to be $\SI{0.20(1)}{\micro\kelvin}$ and $\SI{0.33(1)}{\micro\kelvin}$ respectively. The measured conductance is in qualitative agreement with a calculation based on the Landauer-B\"{u}ttiker formula and the solution of the one-dimensional Schr\"odinger equation for our potential (Appendix \ref{A:Non-interacting simulations}), shown in Fig.~\ref{fig:site_number}(b).

The evolution of transport with the number of barriers represents a direct investigation of the scaling of conductance with the system size. The dependence of $G$ on the barrier number $N$ is presented in Fig.~\ref{fig:site_number}(c) for several values of $V_g$.
Over the whole length range $N = 1, ..., 9$ the evolution of $G$ with number of barriers is well fitted by an exponential decay.
We first use these fits to estimate the initial variation of conductance with length, yielding $\frac{dG}{dN}$ at $N = 1$ as a function of $V_g$ [Fig.~\ref{fig:site_number}(d)].
A sharp threshold at $V_{g,c} = \SI{0.85(2)}{\micro\kelvin}$ is observed: below $V_{g,c}$, conductance is unaffected by extending the lattice beyond a single barrier --- a regime that can be empirically termed as `ballistic'. On the contrary, we measure above $V_{g,c}$ a conductance decay of $1/h$ per additional lattice site --- a `non-ballistic' regime.

The fit also provides a decay coefficient $\kappa$ which, for single particles with energies inside the band gap, should reflect the non-zero imaginary part of their Bloch wavevector in an infinite lattice. We present in Fig.~\ref{fig:site_number}(e) the associated length scale $\kappa^{-1}$ as a function of $V_g$.
Inside the band gap for $V_g \approx \SI{1.15}{\micro\kelvin}$ (white area), it is about two lattice periods $d$ and is indeed comparable to the minimum decay length $1.4 \, d$ that is theoretically expected for Bloch waves.
For gate potentials $V_g < \SI{1.05}{\micro\kelvin}$ inside the band, the precise evolution of conductance with number of barriers greatly depends on the uniformity of the potential, as observed with numerical simulations in Appendix~\ref{A:Non-interacting simulations}. We therefore do not give any specific meaning to the minimum measured at $V_g \approx V_{g,c}$. For $V_g > \SI{1.3}{\micro\kelvin}$ above the gap, the second transverse mode of the wire then becomes populated and conductance cannot be interpreted as the transmission of a one-dimensional lattice.

The decay length is overall bounded by the length of the quantum wire, set by the shorter confining beam with a $1/e^2$ waist of $\SI{9}{\micro\meter}$, which is about 9 lattice periods $d$. It is below $6 \, d$ for most values of $V_g$, which motivates the use of a finite-size lattice made of 6 barriers to investigate the properties of the infinite system in what follows.

\section{Conductor-insulator transition} \label{sec:condins}

\begin{figure}
    \includegraphics{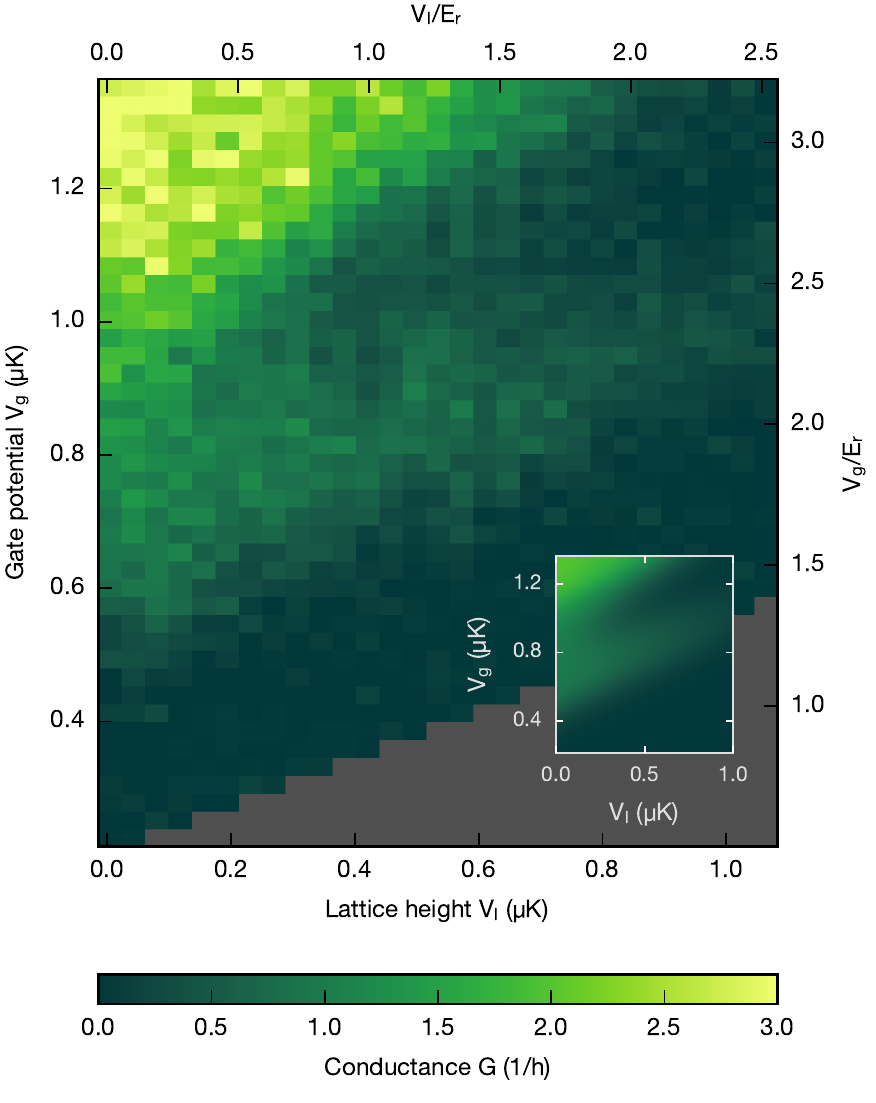}
    \caption{Opening a gap by increasing the lattice height. Experimental conductance through a 6-barrier lattice as a function of lattice height $V_l$ and gate potential $V_g$ (normalized by the recoil energy $E_r = \SI{0.42}{\micro\kelvin}$ in the right and top axes). Two conduction regions (tapered zones in light green and yellow) are separated by an insulating region that broadens upon increasing the lattice height. Inset: Conductance obtained from Landauer-B\"{u}ttiker theory as a function of lattice height $V_l$ and mean chemical potential in the reservoirs $\mu_\text{res}$ at a temperature $T = \SI{60}{\nano\kelvin}$ through a realistic 6-barrier lattice.}
    \label{fig:height}
\end{figure}

We now map out the conductance of a 6-barrier lattice as a function of both chemical potential and lattice depth, demonstrating the emergence of a band structure in a different way. The full map is presented in Fig.~\ref{fig:height}. For low attractive gate potential $V_g$, the lattice is empty and the conductance is zero. Upon increasing $V_g$, the lattice band is visible as a first triangular lobe of non-zero conductance. Its bandwidth decreases from $\SI{0.4}{\micro\kelvin}$ to less than $\SI{0.1}{\micro\kelvin}$ by increasing the lattice height $V_l$ from $0.2$ to $\SI{1.0}{\micro\kelvin}$, and the band is shifted upwards of as a result of a larger lattice zero-point energy. A second triangular lobe of even larger conductance, associated with additional transport in the second transverse mode of the wire, is visible above the band. Both lobes are separated by a gap that increases with $V_l$. The experimental data are in very good agreement with finite-temperature, non-interacting theory shown as inset in Fig.~\ref{fig:height} except in the two-mode regime at large $V_g$, where conductance is larger than two conductance quanta $2/h$. This excess, already visible in Fig.~\ref{fig:site_number}(a) and previously observed in \cite{krinner_mapping_2016}, can be attributed to the presence of attractive interactions.

\begin{figure}
    \includegraphics{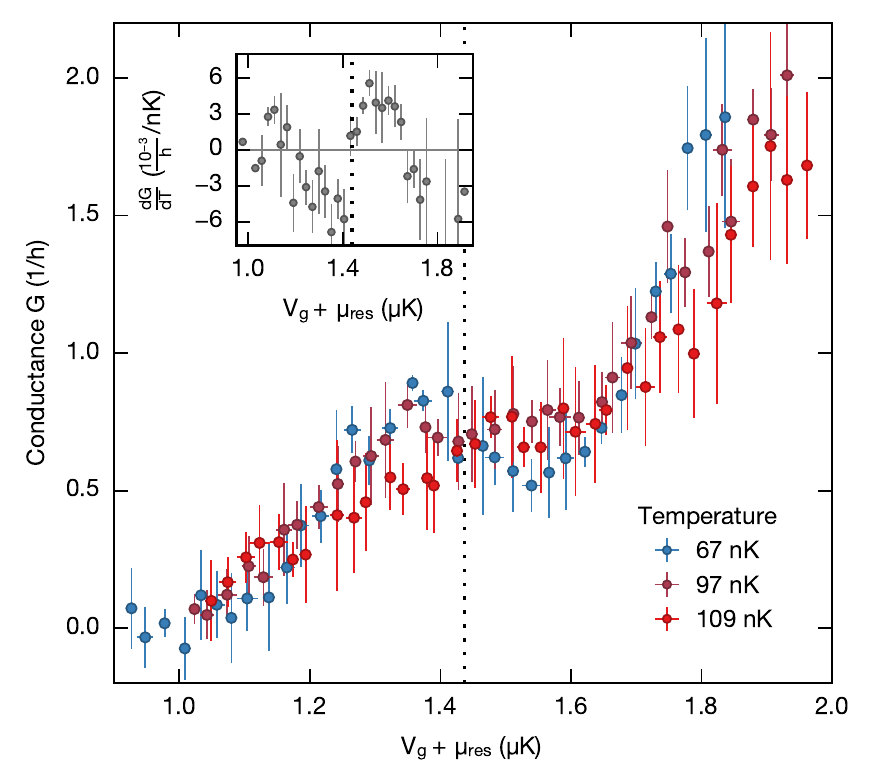}
    \caption{Smearing conductance by thermal decoherence. Experimental conductance $G$ through a 6-barrier lattice of height $V_l = \SI{0.46(2)}{\micro\kelvin}$ for three different temperatures as a function of gate potential $V_g$, shifted by the reservoir chemical potential $\mu_{\text{res}}$ to account for different optical trap frequencies. Inset: Variation of conductance with temperature obtained from a linear fit to temperatures $T = 67, 84, 97$ and $\SI{109}{nK}$. The sign of $dG/dT$ changes across the critical value $V_g + \mu_{\text{res}} = \SI{1.44}{\micro\kelvin}$.}
    \label{fig:temperature}
\end{figure}

The experimental data shown in Fig.~\ref{fig:height} provide an estimate of the centers of the band and gap, visible as local maxima and minima of conductance. To locate more precisely the band conductor-to-insulator transition, we monitor the variation of conductance with increasing temperature, which is positive for an insulator and negative for a conductor. For this purpose, we use a 6-barrier lattice of height $V_l = \SI{0.46(2)}{\micro\kelvin}$, and perform an adiabatic compression of the reservoirs in the transverse direction, in order to vary temperature from $\SI{67}{\nano\kelvin}$ to $\SI{109}{\nano\kelvin}$. This process also changes the chemical potential of the reservoir $\mu_{\rm res}$, hence the variation of local chemical potential in the wire is no longer given by the variation of $V_g$ alone but by the variation of $\mu_{\rm res} + V_g$. Fig.~\ref{fig:temperature} shows conductance curves as a function of the gate potential $V_g$ corrected by the chemical potential variation, such that the band and gap positions can directly be compared. The local maximum and minimum are clearly visible at the lowest temperatures, and are blurred into a monotonically increasing curve at $T = \SI{109}{\nano\kelvin}$. This also highlights the role played by finite temperature inside the gap in Fig.~\ref{fig:site_number}, where conductance is non-zero.
Within our measurement accuracy, the curves intersect at $V_g + \mu_{\text{res}} = \SI{1.44}{\micro\kelvin}$, which separates a region where conductance decreases with temperature, $dG/dT < 0$, from a region where conductance increases, $dG/dT > 0$. This variation $dG/dT$ can be accessed from a linear fit on the conductance and is shown as an inset. The point where $dG/dT = 0$ agrees with non-interacting theory and differs from the transition between the ballistic and non-ballistic regimes studied in Fig.~\ref{fig:site_number} (which occurs there at a corrected value $V_{g,c} + \mu_{\text{res}} = \SI{1.32(2)}{\micro\kelvin}$). It is little sensitive to the details of the one-dimensional lattice potential (Appendix~\ref{A:Non-interacting simulations}) and is therefore a more faithful estimation of the band-to-gap boundary.
While the temperature-dependence of conductance is the traditional definition of an insulator adopted in condensed-matter physics, the existence of a conductance minimum is a more practical criterion in our cold-atom realization and the results of Figs.~\ref{fig:site_number} and \ref{fig:temperature} show the intimate link between the two definitions.
In the following, we extend the use of the latter definition to stronger interactions and look for the presence of a conductance minimum as a signature of an insulating state.

\section{Interactions} \label{sec:interactions}

Interferences, giving rise to the band structure, are essentially single-particle properties. The control over interactions in our system offers a unique opportunity to explore the interplay of interferences with interactions.
We now investigate their effect on transport by increasing \emph{attractive} interactions up to the unitary limit. We tune them to scattering lengths $a < -7.5 \cdot 10^3 a_0$ so that the region close to the wire ends is always superfluid for attractive gate potentials $V_g > \SI{0.7}{\micro\kelvin}$, contrary to the measurements shown in Fig.~\ref{fig:site_number}, Fig.~\ref{fig:height} and Fig.~\ref{fig:temperature}.
As in the previous sections, conductance is obtained assuming a linear relation between current and chemical potential bias, which is here small compared to all other energy scales except temperature (Appendix \ref{A:Data analysis}). This assumption is further justified by the absence of non-linearities without lattice at moderate interactions \cite{krinner_mapping_2016}. As the currents measured through the lattice are low, we could not see any conclusive deviation from this hypothesis within our experimental error.

\begin{figure}
    \includegraphics{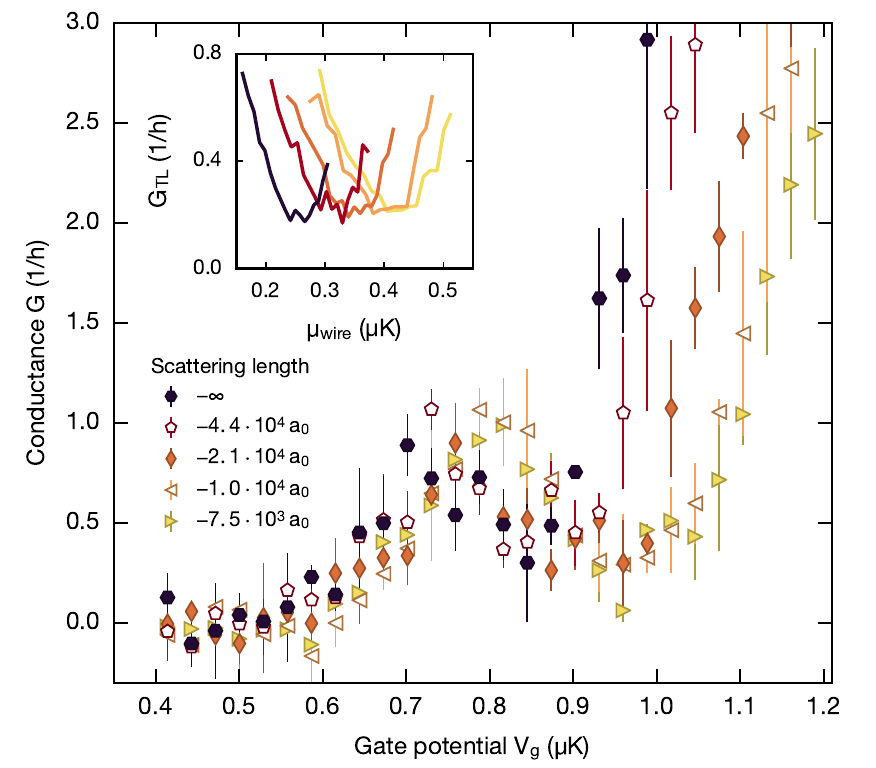}
    \caption{Robustness of the gap with increasing interactions. Conductance $G$ through a 6-barrier lattice of height $V_l = \SI{0.46(2)}{\micro\kelvin}$ as a function of gate potential $V_g$ for different scattering lengths ranging from moderately attractive (yellow) to resonant interactions (dark purple). For a one-dimensional density of two particles per site, where the conductance minimum is observed, the Gaudin-Yang parameter at the wire center is equal to $-1.7, -1.9, -2.3, -2.6$ and $-3.0$ for the five values of the scattering length ranging from $-7.5 \cdot 10^3 a_0$ to $-\infty$; it increases upon decreasing density or equivalently gate potential (Appendix~\ref{A:Bosonization}). Inset: Conductance $G_\text{TL}$ obtained from Tomonaga-Luttinger theory around the conductance dip at temperature $T = \SI{150}{\nano\kelvin}$.}
    \label{fig:interactions}
\end{figure}

The non-monotonic behavior of the conductance versus gate potential is very robust against interactions, as shown in Fig.~\ref{fig:interactions} for a 6-barrier lattice of height $V_l = \SI{0.46(2)}{\micro\kelvin} = 1.09(5) E_r$. The contrast of the gap varies very little as interactions are increased up to unitarity, with a local conductance minimum (resp. maximum) of about $0.3/h$ (resp. $0.9/h$). The positions of the local extrema are shifted towards lower $V_g$ as attractive interactions are increased, which is consistent with an increase of the atom density at the center of the cloud when approaching unitarity.
The persistence of local extrema is very surprising considering the dramatic consequences of attractive interactions on transport in an atomic quantum wire, both experimentally observed and theoretically expected in the absence of a lattice \cite{krinner_mapping_2016,Kanasz-Nagy2016,uchino_anomalous_2017,liu_anomalous_2017}.

The way how the interactions affect transport in a periodic lattice depends on dimensionality and on whether interactions are attractive or repulsive. 
In dimensions greater or equal to two, \emph{repulsive} interactions usually lead to a Fermi-liquid state \cite{nozieres_1999}, very similar to non-interacting particles up to the redefinition of a few parameters such as mass or compressibility. For moderate lattices, the band behavior is thus essentially unaffected. New effects, such as the existence of a Mott insulator for commensurate fillings of one particle per site, can thus only occur for large lattices in the repulsive case.
\emph{Attractive} interactions on the other hand lead to a drastic change of the excitations and turn a fermionic system into a superconductor, 
a collective state for which the flow of particles is impervious to obstacles such as disorder or a weak lattice. 
As a result, the effects of the lattice are reduced and even at a commensurate filling the band gap disappears when it is smaller than the superconducting gap in the absence of a lattice \cite{nozieres_semiconductors_1999}.

This competition between a band insulator and superconductivity has however a very different outcome in one dimension where the effects of interactions are drastically enhanced. These effects are captured at low energy by the Tomonaga-Luttinger liquid theory \cite{giamarchi_quantum_2003}, in which no individual quasi-particles similar to free particles exist, and where excitations of the many-body state separate into collective charge and spin excitations. This has several consequences on the ground state of an interacting quantum system.
For bosons, in marked contrast to the higher-dimensional counterparts, even an \emph{infinitesimal} lattice is able to lead to a Mott state for repulsive interactions at commensurability of one boson per site \cite{haldane_effective_1981}, provided the repulsion between the bosons exceeds a certain threshold. This existence of this critical value and the corresponding phase transition has been demonstrated with cold atoms \cite{buchler_commensurate-incommensurate_2003,haller_pinning_2010,boeris_mott_2016}.
For fermions, repulsive interactions lead to a Tomonaga-Luttinger liquid state \cite{giamarchi_quantum_2003} in which the spin sector has dominant antiferromagnetic tendencies. The charge sector is fully decoupled, spin-charge separation being one of the remarkable properties of one-dimensional interacting 
systems. In a one dimensional periodic structure, the system will turn to be a Mott insulator for a filling of one particle per site for arbitrarily weak repulsion. For two particles per site one recovers a band insulator even in the presence of repulsion between particles. 
For attractive interactions, the outcome is quite different: the spin excitations are gapped, leading to a state where only charge excitations exist at low energy.
This state, known as a Luther-Emery liquid \cite{luther_backward_1974,giamarchi_quantum_2003}, can also be deeply affected even by a very weak lattice contrarily to its high-dimensional BCS counterpart \cite{nozieres_semiconductors_1999}. It can lead to an insulating state with one pair per lattice site on average.
Thus at commensurate filling, even for \emph{infinitely attractive} interactions, conduction is not recovered and the system remains an insulator. This state can then be seen as the many-body equivalent of the band insulator for non-interacting particles, with fillings of two particles per site [Fig.~\ref{fig:setup}(c)].
It is furthermore associated with the existence of charge-density waves, whose observation was proposed through structure factor measurements with cold atoms \cite{xianlong_luther-emery_2007}.

In order to analyze the behavior observed experimentally, we consider a theoretical model of fermionic atoms with an attractive contact interaction in a one-dimensional wire and an external potential. The system is described by the Hamiltonian
\begin{equation}
 H = H_\mathrm{GY} + H_\mathrm{lattice},
 \end{equation}
where the fermions without external potential obey the Gaudin-Yang Hamiltonian \cite{Gaudin1967,Yang1967}
\begin{equation}
    H_\mathrm{GY} = -\dfrac{\hbar^2}{2m} \sum_i \dfrac{\partial^2}{\partial y_i^2} + g_1 \sum_{i<j} \delta(y_i - y_j),\label{eq:Gaudin-Yang}
\end{equation}
where $y_i$ is the position of the $i$-th atom in the wire and $g_1$ is the strength of the short-range interaction.
The influence of the optical lattice is taken into account by
\begin{equation}
H_\mathrm{lattice} = \int dy\, V(y)\rho(y)\label{eq:Gaudin-Yang2},
\end{equation}
where $V(y)$ is the potential of the lattice and $\rho(y)$ is the total local density of fermions.

We proceed by treating the experimentally relevant low-energy degrees of freedom of the Gaudin-Yang model \eqref{eq:Gaudin-Yang} with the Tomonaga-Luttinger liquid theory \cite{giamarchi_quantum_2003, cazalilla_one_2011}. This provides the description of the gap in the spin sector, and the transformation
to the Luther-Emery liquid made of bound pairs of finite extent, strongly repelling each other as a result of the exclusion principle between their fermionic constituents. The resulting system is described by a sine-Gordon equation, whose parameters can be obtained as a function of the strength of the short-range interaction (Appendix~\ref{A:Bosonization}).

We then compute the transport properties of the system by attaching this one-dimensional system to two reservoirs. The choice of a one-dimensional model is justified here by the fact that the $\SI{9}{\micro\meter}$-wire is longer than the superfluid coherence length $\hbar v_F/ T \approx \SI{3}{\micro\meter}$ in the reservoirs, where $v_F$ is the Fermi velocity. This situation is different from previous works with a short quantum point contact \cite{husmann_connecting_2015} where the physics is governed by the reservoir-induced proximity effect. We neglect the contact resistance compared to the resistance of the scattering potential in the wire, and use the approximation of one-dimensional leads \cite{Maslov1995}.

We evaluate the conductance by numerically solving the sine-Gordon equation mentioned above in a noisy thermal background (Appendix~\ref{A:Conductance}).
The results are shown as inset in Fig.~\ref{fig:interactions} as a function of the chemical potential in the wire $\mu_\text{wire} \approx V_g - \mathrm{const.}$, where $V_g$ is the gate potential and the constant contribution is due to the chemical potential of the reservoirs at the wire entrance. They are in good qualitative agreement with the experiment and correctly predict a conductance minimum compatible with the formation of an insulator.
This occurs at lattice fillings of two fermions per site, which can be translated into local chemical potentials $\mu_\text{wire} = 0.25$ to $\SI{0.45}{\micro\kelvin}$ using an approximate equation of state for the wire (Appendix~\ref{A:Conductance}). A better quantitative agreement for the value of the conductance is found by increasing the effective temperature used in the simulation ($\SI{150}{\nano\kelvin}$ in the inset of Fig.~\ref{fig:interactions}, compared to about $\SI{70}{\nano\kelvin}$ in the experiment).
This discrepancy may stem from neglecting the influence of the reservoirs and from taking into account only classical fluctuations of the bosonized fields.

The robustness of the insulating state even at unitarity strongly indicates that we indeed realize the Luther-Emery state inside the wire.
An alternative way to understand the Luther-Emery liquid is to consider a one-dimensional theory where the elementary constituents are not the fermionic atoms, but instead weakly-bound bosonic pairs with an effective finite-range repulsion (Appendix~\ref{A:Super-Tonks-Girardeau gas}). These pairs form a so-called super-Tonks-Girardeau gas (STG). The insulating state can be identified with a Mott-type insulator of bosons \cite{giamarchi_quantum_2003, cazalilla_one_2011}.  We emphasize that in contrast to previous works, where STG gases were theoretically predicted \cite{Astrakharchik2005} and experimentally realized \cite{Haller2009} as a highly-excited and strongly-correlated metastable gas-like state of \emph{attractive} bosons, in our case the STG phase is realized with spin-half fermions \cite{chen_realization_2010}. The finite size of the pairs, which is a key ingredient to the finite-range repulsion, allows us to obtain the essential properties of the STG gas as a stable ground state.
This demonstrates the potential of our fermionic setup to simulate novel one-dimensional bosonic phases as well.

\section{Conclusions} \label{sec:conclusions}

In this work, we demonstrated local control of the potential landscape in a quantum wire to study the transport properties of a one-dimensional fermionic insulator created by a mesoscopic lattice, in the presence of attractive interactions.
We were able to observe the effect of quantum interferences on transport by projecting one by one optical barriers varying on distances of the order of the Fermi wavelength. By mapping out conductance against chemical potential, lattice height and temperature, we confirm that our observations are in agreement with the physics of a band insulator and confront our system with the traditional definition of an insulating phase.

These results serve as a reference point to study the fate of the band insulator with stronger attractive interactions, where the superfluid gap in the wire is expected to be larger than the band gap. Surprisingly, the system remains insulating at commensurate filling for arbitrarily large attractive interactions. We explain this peculiar behavior by the formation of a so-called Luther-Emery liquid. The conductance measurements are qualitatively reproduced by a Tomonaga-Luttinger model, allowed by the high control offered by the experimental setup. Our experiment thus constitutes a direct observation of the existence of the Luther-Emery phase in a one-dimensional system of fermions with attractive interactions.

The system demonstrate the possibility to realize transport measurements between two reservoirs of fermions with tunable interactions and an arbitrary potential, allowing us to probe a very rich palette of physical effects.
Beside an extended study of the Luther-Emery phase as a function of lattice height and temperature, the effect of the competition between potential and interactions on transport could be investigated when one or a few impurities are added to a Tomonaga-Luttinger liquid, or in presence of disorder. There, size represents a controllable length scale which is crucial to determine the physical nature of the many-body system. For example, one expects a linear scaling between conductance and system size for normal metals, and an exponential scaling for insulators (such as e.g.\ an Anderson or many-body localized insulator).

The structures projected onto the transport channel can be readily extended to two-dimensional lattices, with the advantage that chemical potential and temperature can be tuned independently of the lattice parameters. This opens the path to investigating edge effects or creating exotic band structures showing an interesting interplay with interactions, such as semi-metals or flat bands. More generally, our system is relevant for studying transport processes where the energy-dependence of the conduction channel plays a crucial role, for example to introduce thermoelectric coupling \cite{grenier_peltier_2014}.

By varying the arbitrary potentials over time, our setup allows us in principle to study the time-dependent response of correlated systems, implement quantum pumps \cite{switkes_adiabatic_1999,nakajima_topological_2016,lohse_thouless_2016}, or Floquet-engineered topological bands \cite{jotzu_experimental_2014}. Close-to-resonance beams could be furthermore used to implement spin-dependent potentials or spin-orbit coupling locally, or to study the effect of dissipation on quantum coherence and superfluidity.

\begin{acknowledgments}
We thank S.~Krinner and S.~Nakajima for experimental assistance at the beginning of the project and R.~Desbuquois, S.~Peotta, P.~T\"{o}rm\"{a} and W.~Zwerger for careful reading of the manuscript and discussions.

We acknowledge support from Swiss NSF under division II (Project Number 200020-162519, 200020-169320 and NCCR-QSIT), Swiss State Secretary for Education, Research and Innovation Contract No. 15.0019 (QUIC), ERC advanced grant TransQ (Project Number 742579) and ARO-MURI Non-equilibrium Many-body Dynamics grant (W911NF-14-1-0003) for funding.
\mbox{L.C.} is supported by ETH Zurich Postdoctoral Fellowship and Marie Curie Actions for People COFUND program.
\mbox{J.-P.B.} is supported by the ERC project DECCA (Project Number 714309) and the Sandoz Family Foundation-Monique de Meuron program for Academic Promotion. 
\end{acknowledgments}

\appendix

\section{Experimental details}
\label{A:Experimental details}

\subsection{Experimental sequence}

The experimental cycle to prepare and manipulate an ultracold Fermi gas of $^6$Li atoms is based on \cite{krinner_observation_2015,krinner_mapping_2016}. In summary, the atom cloud is prepared in a balanced mixture of the lowest and third lowest hyperfine states and loaded into an elongated hybrid trap, confining transversally ($x, z$ directions) with a $\SI{1064}{\nano\meter}$ optical dipole trap and longitudinally ($y$ direction) by a quadratic variation of the magnetic field.
Shortly before performing a conductance measurement, the gas is brought to temperatures close to $10 \%$ of the Fermi temperature through evaporative cooling on a broad Feshbach resonance at 689 G.
Its absolute temperature $T$ is fixed by the confinement frequencies of the dipole trap, which are directly related to its depth, set to $\SI{1.50}{\micro\kelvin}$ in Figs.~\ref{fig:site_number}, \ref{fig:height} and \ref{fig:interactions} ($T = \SI{67}{\nano\kelvin}$), and between 1.50 and $\SI{4.51}{\micro\kelvin}$ in Fig.~\ref{fig:temperature} ($T = 67 - \SI{109}{\nano\kelvin}$).
To set the s-wave scattering length $a$ for transport measurements, the magnetic field is then ramped in $\SI{200}{\milli\second}$ to a value of $\SI{949}{G}$ in Figs.~\ref{fig:site_number}, \ref{fig:height} and \ref{fig:interactions} ($a = -2.65 \cdot 10^3 \,a_0$), and between 689 and $\SI{726}{G}$ in Fig.~\ref{fig:temperature} ($a = -\infty$ to $-7.51 \cdot 10^3 \,a_0$).

To impose a chemical potential bias, we displace the cloud using a magnetic field gradient in the $y$ direction. We then split it into two asymmetric reservoirs using an elliptical repulsive beam at $\SI{532}{\nano\meter}$, superimposed to the laser beams shaping the quantum wire and the lattice.
The magnetic trap is then recentered on the constriction beams.
The atom flow through the constriction is initiated by switching off the elliptical beam and lasts for 4 s.
Subsequently, the magnetic field is ramped within $\SI{200}{\milli\second}$ back to the Feshbach resonance, which is where we finally obtain the density distribution by absorption imaging along the $x$ direction after a time-of-flight of $\SI{1}{\milli\second}$.

\subsection{Holographic beam shaping}

The lattice is shaped using a Digital Micromirror Device (DMD DLP5500 .55" XGA from Texas Instruments), consisting of an array of microscopic mirrors that can be individually oriented to an ON or an OFF state and act altogether as a reflective diffraction grating.
We illuminate the DMD with a collimated, $\SI{532}{\nano\meter}$ beam with an incidence angle that allow a 30 \% diffraction efficiency into the $6^\text{th}$ order when all mirrors are ON.
The hologram represented on the array is a binary mask of lines whose local width and displacement affect the local amplitude and phase of the diffracted beam.
This beam is then optically conjugated using relay lenses to the back focal plane of a microscope objective, which effectively projects its Fourier transform onto the atomic plane.
A second microscope objective placed symmetrically with respect to the atomic plane allows us to image the light potentials created by the DMD on a CCD camera.
In addition, we can correct optical aberrations to limit the wavefront distortion on the atoms to about $\lambda/10$ using a technique similar to a Hartmann-Shack analysis.

To project the lattice onto the quantum wire, we first align the relevant DMD diffraction order by scanning it across the wire and measuring the subsequent variations of atom current, a technique known as scanning gate microscopy \cite{hausler_scanning_2017}.
The DMD beam is located on the wire center when the conductance is minimal.
The lattice hologram itself is displayed on a grid of 660 by 660 mirrors.
The size of the hologram and the number of ON mirrors decreases when the lattice is made longer in the associated Fourier plane. Hence, we observe a drop of the intensity diffracted from the DMD, from 13 \% of the incident intensity for a single barrier to 0.8 \% for a 10-barrier lattice.
The potential height at the center of the lattice is calibrated for all lengths to account for this drop.

\section{Data analysis}
\label{A:Data analysis}

\subsection{Energy scales}

\begin{figure}
    \includegraphics{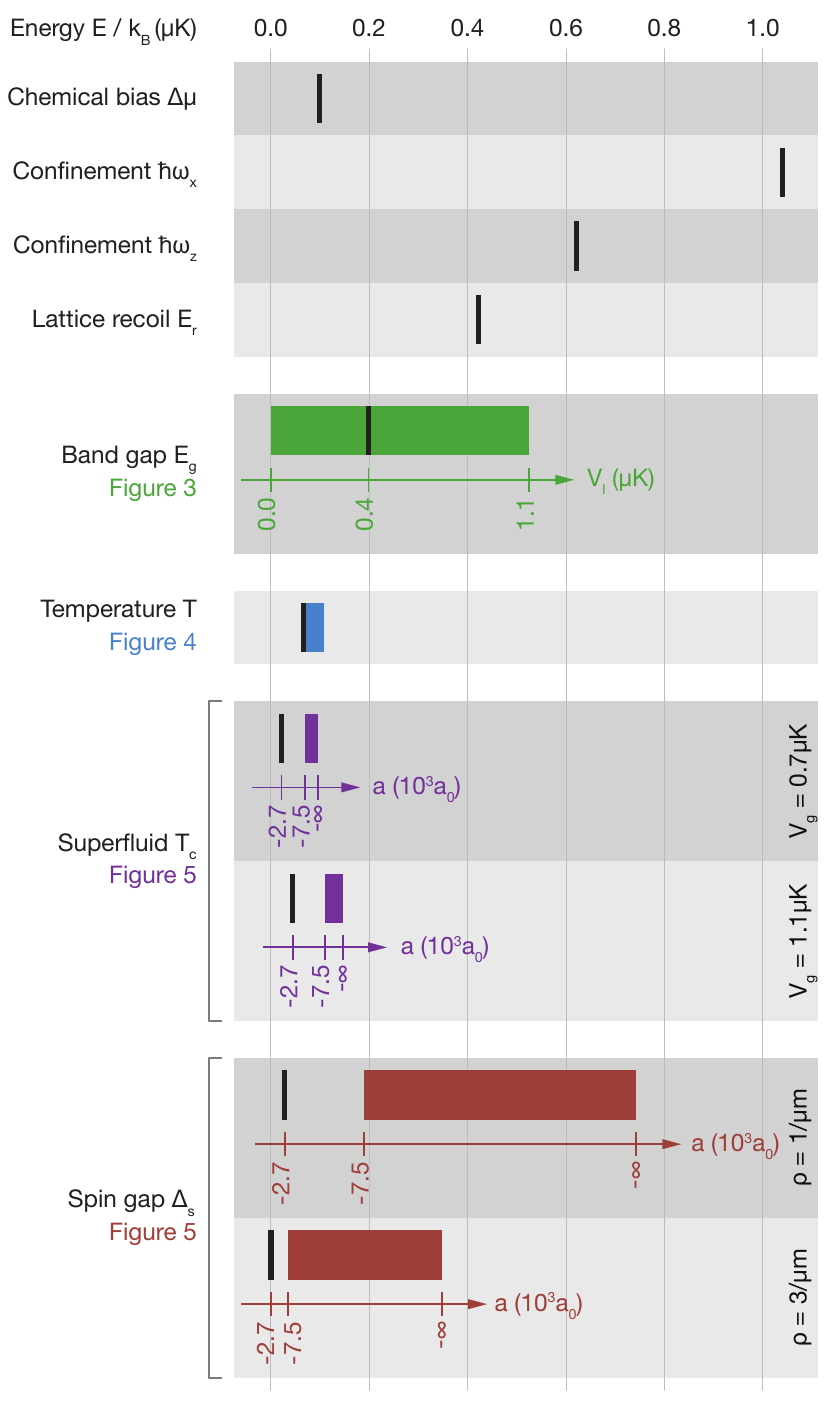}
    \caption{Summary of the energy and temperature scales relevant for the quantum wire. Typical values are indicated by black lines, whereas energies depending on a parameter scanned in Figs.~\ref{fig:height}, \ref{fig:temperature} and \ref{fig:interactions} are represented by colored ranges. The superfluid critical temperature $T_c$ in the vicinity of the wire increases with gate potential $V_g$, and is reported here for two values $V_g = \SI{0.7}{\micro\kelvin}$ and $\SI{1.1}{\micro\kelvin}$ corresponding to the conductance onset and gap with a lattice respectively. The 1D spin gap $\Delta_s$ at the center of the wire decreases with the 1D density $\rho$, and is reported here for two values corresponding to a lattice filling of one and three particles per site.}
    \label{fig:energy_scales}
\end{figure}

The energy and temperature scales relevant in the experiment are summarized in Fig.~\ref{fig:energy_scales}.

\begin{itemize}
\item The chemical potential bias $\Delta \mu = \SI{0.1}{\micro\kelvin}$ is smaller than the energy scales characterizing the wire, lattice and spin gap. This supports the use of a linear model to obtain the conductance.
\item The confinement energies $\hbar \omega_x$ and $\hbar \omega_z$ are the largest energy scales, confirming that our system is effectively one-dimensional (1D) when the local chemical potential is sufficiently low.
\item The lattice recoil energy $E_r = \SI{0.42}{\micro\kelvin}$ is comparable to the typical lattice height $V_l = \SI{0.4}{\micro\kelvin}$, placing the system in the nearly free particle regime rather than the tight-binding regime.
\item The band gap $E_g = \SI{0.2}{\micro\kelvin}$ is larger than both temperature $T$ and chemical potential bias $\Delta \mu$ in a typical setting. It becomes unresolved in conductance measurements when the lattice height $V_l$ is decreased (Fig.~\ref{fig:height}) or the temperature is increased (Fig.~\ref{fig:temperature}).
\item At the lowest scattering length $a = -2.65 \cdot 10^3 \, a_0$ used in the experiment, temperature $T$ is larger than both the critical superfluid temperature $T_c$ in the vicinity of the wire and the spin gap $\Delta_s$ in the wire, indicating that the system is in the normal phase.
\item On the contrary, for scattering lengths used in Fig.~\ref{fig:interactions}, the regions of the Fermi gas connected to the wire are superfluid ($T < T_c$). The spin gap $\Delta_s$ is larger than temperature except in the case of the smallest $a$ and largest densities, which supports the Luther-Emery liquid hypothesis. Depending on interaction and density, $\Delta_s$ may be smaller (small $a$/large $\rho$) or larger (large $a$/small $\rho$) than the lattice gap $E_g$. Mean-field theory predicts the presence of a transition in 2D and 3D when $E_g \approx \Delta_s$ \cite{nozieres_semiconductors_1999}, which is here not observed experimentally.

\end{itemize}

\subsection{Temperature estimation}

For Figs.~\ref{fig:site_number} to \ref{fig:temperature}, temperature is extracted from the density pictures of the cloud in the magnetic harmonic trap along $y$ taken at unitarity.
To filter out imaging noise, we apply a principal component analysis \cite{Segal2010} on sets of pictures belonging to the same experimental run grouped by trap depth.
This method reconstructs each picture from the mean picture of the series and a linear combination of the five most relevant deviations to the mean.
We obtain the internal energy from the second moment along $y$ using the virial theorem \cite{Thomas2005}, which is then converted into entropy using the known equation of state of the unitary Fermi gas \cite{Ku2012}, approximating the hybrid trap by a harmonic potential.
Assuming that the magnetic field and trap depth ramp preceding imaging is adiabatic, we equate the entropies of the gas at imaging and at the end of transport, where it can be expanded for the weakly interacting, degenerate gas \cite{Su2003}
\begin{equation}
\frac{S}{N} = k_B \pi^2 \frac{T}{T_F} \left(1 + \frac{64}{35\pi^2} k_F a \right),
\end{equation}
where the Fermi temperature $T_F = \hbar \bar{\omega} (6 N)^{1/3}/k_B$ is inferred from the atom number $N$ and the mean trapping frequency $\bar{\omega} = (\omega_x \omega_y \omega_z)^{1/3}$, and the corresponding Fermi wavevector is defined as $k_F = \sqrt{2 m k_B T_F}/\hbar$.
This allows us to access the temperature $T$ at the end of transport.
The temperatures stated in the text are estimated at half the transport time, and are obtained after subtracting the heating occurring during $\SI{2}{\second}$ in the dipole trap.
This heating rate is independently calibrated and ranges from $\SI{1(2)}{\nano\kelvin/s}$ (Figs.~\ref{fig:site_number}, \ref{fig:height} and lowest temperature of Fig.~\ref{fig:temperature}), to $\SI{8(2)}{\nano\kelvin/s}$ (largest temperature of Fig.~\ref{fig:temperature}).

The superfluid critical temperatures stated in the text and shown in Fig.~\ref{fig:energy_scales} are computed at the minima of the quasi-one-dimensional potential located at the entrance and exit of the quantum wire \cite{krinner_mapping_2016}. Under the assumption that the local chemical potential $\tilde{\mu}$ at these points is about the Fermi energy $\tilde{E}_F$ in a local density approximation, we compute a local interaction parameter $1/\tilde{k}_F a$, which is then converted into a critical temperature using \cite{haussmann_thermodynamics_2007}.

\subsection{Thermodynamical properties of the reservoirs}

In Figs.~\ref{fig:site_number} to \ref{fig:temperature}, we use the equation of state of the homogeneous, weakly-interacting Fermi gas \cite{Su2003} to compute the density $\tilde{\rho}(\vec{r})$ for one spin species in the local density approximation:
\begin{equation}
\tilde{\rho}(\vec{r}) = -\frac{1}{\lambda_T^3} \text{Li}_{3/2}(-\tilde{z})\left(1+\frac{2 a}{\lambda_T} \text{Li}_{1/2}(-\tilde{z})\right)
\end{equation}
where $\tilde{z}(\vec{r}) = \exp[\tilde{\mu}(\vec{r})/k_B T]$ is the local fugacity of the gas, $\tilde{\mu}(\vec{r}) = \mu_\text{res}-V(\vec{r})$ is the local chemical potential in the known potential $V(\vec{r})$ created by the trap and the constriction, $\lambda_T = \sqrt{2 \pi \hbar^2/m k_B T}$ is the thermal de Broglie wavelength and $\text{Li}_n$ indicates the polylogarithm function of order $n$. The local compressibility can be inferred from the density as $\tilde{\kappa} = \partial \tilde{\rho}/\partial \tilde{\mu} \rvert_a$.

The atom number $N$ and compressibility $\kappa$ of the entire reservoirs are then obtained by spatially integrating $\tilde{\rho}(\vec{r}), \tilde{\kappa}(\vec{r})$ and are sampled at fixed temperature $T$ for several values of the chemical potential in the reservoirs $\mu_\text{res}$. The interpolated functions $\mu_\text{res}(N), \kappa(N)$ are then used to estimate the chemical potential in each reservoir and the mean compressibility for every absorption picture. Within linear approximation, conductance can be inferred from the atom number difference between the reservoirs after 4 s of transport and the compressibility \cite{krinner_observation_2015}.

To account for the wide range of interactions in Fig.~\ref{fig:interactions}, we instead use there the zero-temperature equation of state of the balanced Fermi gas across the BEC-BCS crossover. The local pressure $\tilde{P}(\tilde\mu)$ created by one spin species as a function of the local chemical potential is
\begin{equation}
\tilde{P}(\tilde\mu) = P_0\,(\tilde\mu) h_S^{\text{BCS}}(\hbar/\sqrt{2 m \tilde\mu}a),\label{eq:P}
\end{equation}
where $P_0 = \pfrac{2m}{\hbar^2}^{3/2} \tilde{\mu}^{5/2}/15\pi^2$ is the pressure of a spin-polarized ideal Fermi gas and $h_S^{\text{BCS}}(\tilde\delta)$ is a correction factor determined experimentally \cite{Navon2010}. Its derivatives yield the local density $\tilde{\rho} = \partial \tilde{P}/\partial \tilde{\mu} \rvert_a$ and compressibility $\tilde{\kappa} = \partial \tilde{\rho}/\partial \tilde{\mu} \rvert_a$. We then apply the procedure described above to obtain compressibility and conductance for every absorption picture.

\begin{figure*}[!htb]
    \includegraphics{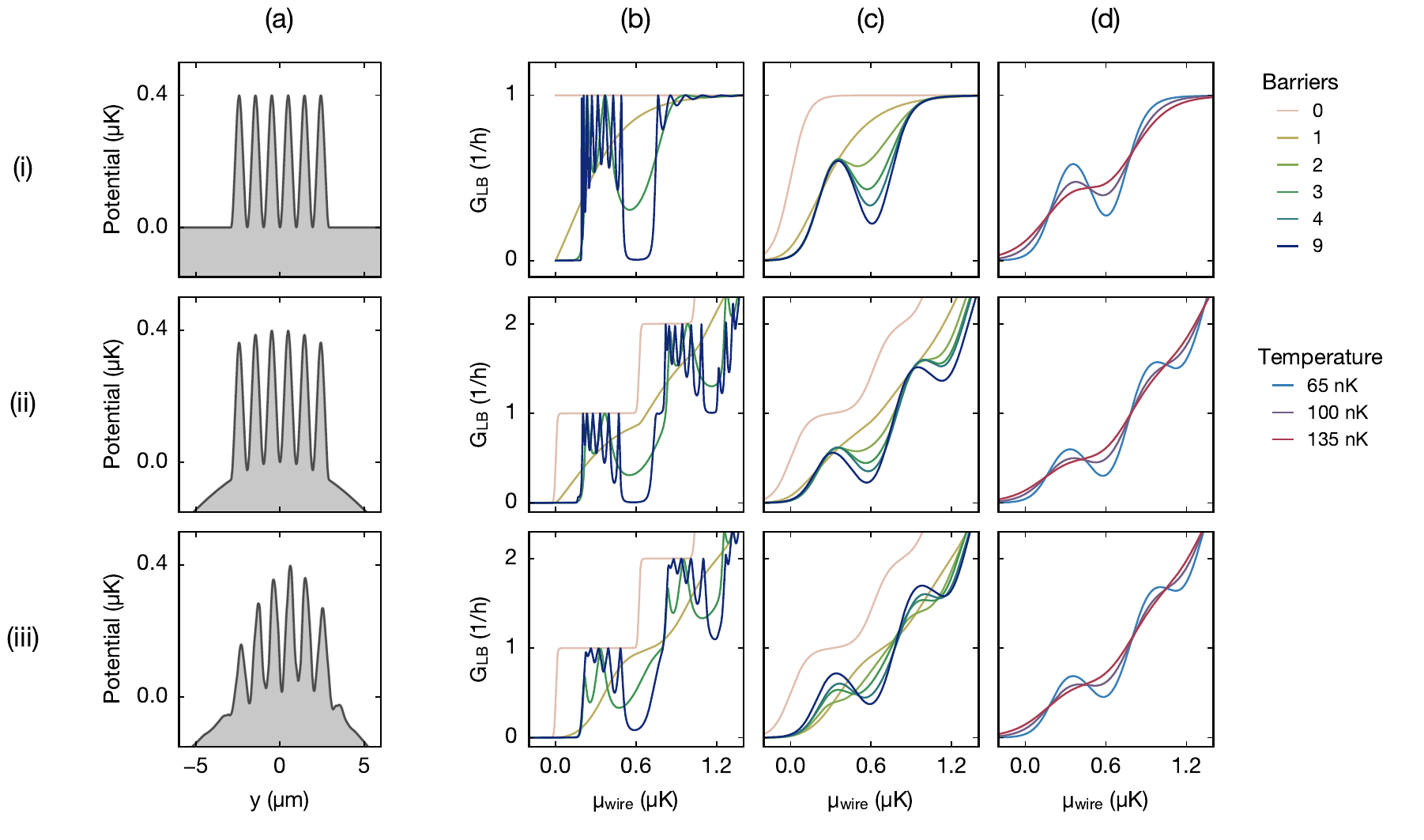}
    \caption{Effect of the lattice potential on conductance, obtained from non-interacting Landauer-B\"{u}ttiker theory. Results for: (i) an ideal one-dimensional lattice potential on a flat background; (ii) an ideal lattice in the quasi-one-dimensional potential created by the quantum wire; (ii) a real lattice in the effective potential of the quantum wire. Column (a): effective potential of the lowest transverse mode. Column (b): conductance $G_\text{LB}$ versus local chemical potential in the wire $\mu_\text{wire}$ for different numbers of barriers $N$ at temperature $T = 0$ and infinitesimal bias $\Delta \mu$. Column (c): id. with $T = \SI{60}{\nano\kelvin}$ and $\Delta \mu = \SI{100}{\nano\kelvin}$. Column (d): conductance $G_\text{LB}$ versus chemical potential $\mu_\text{wire}$ for $N = 6$ barriers, $\Delta \mu = \SI{100}{\nano\kelvin}$ and different temperatures $T$.}
    \label{fig:numerov}
\end{figure*}

\section{Non-interacting simulations}
\label{A:Non-interacting simulations}

\subsection{Landauer-B\"{u}ttiker formula}

To model the experimental results at the lowest scattering length $a = -2.65 \cdot 10^3 \, a_0$ presented in Figs.~\ref{fig:site_number} to \ref{fig:temperature} and include the presence of higher-energy modes above the transverse ground state of the constriction, we apply the Landauer-B\"{u}ttiker formalism \cite{Ihn2009}. It separates the conductance of non-interacting fermions into contributions from independent transport channels:
\begin{equation}
G = \frac{1}{h \cdot \Delta \mu} \sum_{n,m} \int_{-\infty}^\infty dE \mathcal{T}_{n, m}(E) [f_L(E) - f_R(E)].
\label{eq:landauer}
\end{equation}
Every channel is associated to a transverse mode of the quantum wire, indexed by the quantum numbers $n, m$ along directions $x$ and $z$ respectively, and characterized by a transmission coefficient $\mathcal{T}_{n,m}(E)$ whose energy-dependence reflects the interference processes at work inside the mesoscopic structure. In this picture, transport results from the difference in the Fermi-Dirac distributions $f_L$, $f_R$, describing the particle energies in the two reservoirs linked by the channels. The function $f_L$ (respectively $f_R$) is temperature-dependent and centered around the chemical potential $\mu_L$ ($\mu_R$) of the left (right) reservoir. The chemical potential bias that drives the atom current is defined as $\Delta \mu = \mu_L - \mu_R$.

The transmission $\mathcal{T}_{n, m}(E)$ of each channel is obtained from the squared scattering amplitude of a single particle through a one-dimensional effective potential $V_{\text{eff}, n, m}(y)$. This potential has contributions from each of the four beams that shape the constriction: the transverse energies of the space-dependent harmonic confinement provided by the wire in directions $x$ and $z$, $V_{x,n}(y) = (1/2+n) \hbar \omega_x f_x(y)$ and $V_{z,m} = (1/2+m) \hbar \omega_z f_z(y)$, the attractive gate potential $V_g(y) = -V_g f_g(y)$ and the lattice potential $V_l(y)$.
The envelope functions $f_x$, $f_y$, $f_z$ incorporate known information on the geometry of these laser beams and are detailed in Table \ref{table:envelope}.
Additionally, the conductance shown in Fig.~\ref{fig:site_number}(b) includes a repulsive potential of $\SI{0.3}{\micro\kelvin}$, necessary to match the $V_g$-axis of the experimental data, and resulting from a probable non-darkness of the repulsive $\text{TEM}_{01}$-like laser beams.

\begin{table}
\begin{tabular}{lll}
  Envelope function & Waist & Description \\
  \hline
  $f_x(y) = \exp(-y^2/w_x^2)$ & $w_x = \SI{9.1}{\micro\meter}$ & $x$ confinement \\
  $f_z(y) = \exp(-y^2/w_z^2)$ & $w_z = \SI{30.2}{\micro\meter}$ & $z$ confinement \\
  $f_g(y) = \exp(-2 y^2/w_g^2)$ & $w_g = \SI{42.5}{\micro\meter}$ & gate potential \\
  \hline
\end{tabular}
\caption{Envelope functions determining the effective quasi-one-dimensional potentials
}
\label{table:envelope}
\end{table}

\subsection{Non-uniformity of the scattering potential}

The effective one-dimensional potential differs from an ideal finite-length lattice due to the presence of the quantum wire effective potential $V_{x,n}(y) + V_{z,m}(y)$ and to the non-uniform envelope the lattice potential $V_l(y)$. Both factors contribute to reducing the height of the outermost parts of the lattice relative to the chemical potential of the reservoirs.

To illustrate these two effects on conductance independently, we simulate the conductance measurements shown in Figs.~\ref{fig:site_number} and \ref{fig:temperature} for different one-dimensional potentials with lattice height $V_l = \SI{0.40}{\micro\kelvin}$ [Fig.~\ref{fig:numerov}(a)]:
\begin{enumerate}[label=(\roman*)]
\item with a sine lattice of $N$ barriers and spacing $d = \SI{0.97}{\micro\metre}  $, on top of a flat background: $V_{\text{eff}, n, m}(y) = V_{l, \text{ideal}}(y) = V_l \sin[\pi(y/d+N/2)]^2$ if $y \in [-N d/2, N d/2]$, and $V_{\text{eff}, n, m}(y) = 0$ else;
\item with a sine lattice added to the effective potential of the quantum wire, $V_{\text{eff}, n, m}(y) = V_{x,n}(y) + V_{z,m}(y) + V_{l, \text{ideal}}(y)$;
\item with a real lattice obtained by directly imaging the optical intensity of the lattice beam, added to the effective potential of the quantum wire, $V_{\text{eff}, n, m}(y) = V_{x,n}(y) + V_{z,m}(y) + V_{l, \text{real}}(y)$.
\end{enumerate}

The transmission amplitude of a free particle incident on the potential is obtained by solving the Schr\"{o}dinger equation numerically over a length which is as long as the lattice for potential (i), and which is $\SI{120}{\nano\metre}$ larger than the $1/e^2$ waists $w_x$ and $w_z$ of the envelopes of the quantum wire for potentials (ii) and (iii).
The sum over the transverse modes $n, m$ in \eqref{eq:landauer} is truncated to the lowest energy channel in case (i), enforcing the maximum conductance to $1/h$ [Fig.~\ref{fig:numerov}(i)--(b) to (d)], and to the five lowest channels in cases (ii) and (iii), giving rise to conductance plateaus at multiples of $1/h$ at zero temperature and without lattice potential [Fig.~\ref{fig:numerov}(ii)--(b) and (c)].
The effect of the attractive gate potential on this region (where it is homogeneous to a good approximation) is to shift the mean value of the chemical potential $\mu_\text{wire} = (\mu_L + \mu_R)/2$ with respect to the zero-point energy of the channel; its variation amounts to a change of origin of the x-axis in Fig.~\ref{fig:numerov}(b) to (d).

We now simulate scattering through these three potentials in different situations relevant to the experiment: first, we vary the number of barriers $N$ making up the lattice, comparing the ideal case at temperature $T = 0$ and infinitesimal chemical potential bias $\Delta \mu$ [Fig.~\ref{fig:numerov}(b)], where formula \eqref{eq:landauer} simplifies to $G = \sum_{n,m} \mathcal{T}_{n, m}(\mu_\text{wire})/h$, to the more realistic situation where $T = \SI{60}{\nano\kelvin}$ and $\Delta \mu = \SI{100}{\nano\kelvin}$ [Fig.~\ref{fig:numerov}(c)]; second, we vary the temperature at fixed lattice length $N = 6$ [Fig.~\ref{fig:numerov}(d)].

For the ideal lattice potential, the gap is visible as an interval around $\mu_\text{wire} = \SI{0.6}{\micro\kelvin}$ where conductance uniformly converges to zero upon increasing $N$ [Fig.~\ref{fig:numerov}(i)--(b)].
The band appears as a succession of $N-1$ conductance peaks that equal the number of hybridized orbitals expected in a tight-binding model.
When including the potential of the quantum wire [Fig.~\ref{fig:numerov}(ii)--(b)] the number of oscillations decreases, which can be interpreted as a decrease of the effective length of the lattice.
The non-uniform lattice potential further reduces the contrast of the band oscillations and the effective gap width [Fig.~\ref{fig:numerov}(iii)--(b)].

Adding a finite chemical potential bias blurs out the band oscillations but the gap subsists as a local conductance minimum between $0.3/h$ and $0.6/h$ depending on the lattice length, as observed in the experiment.
As the lattice length is increased, the precise evolution of the conductance curves is however sensitive to the details of the potential.
On the contrary, increasing the temperature above $T = \SI{60}{\nano\kelvin}$ at fixed lattice length $N = 6$ [Fig.~\ref{fig:numerov}(d)] softens the local extrema of conductance rather independently of the precise lattice potential.
Importantly, all conductance curves nearly intersect at the chemical potentials demarcating the zero-temperature bands and gaps, justifying the use of $dG/dT$ as an indication of the conductor-insulator transition in Fig.~\ref{fig:temperature}.

\section{Bosonization}
\label{A:Bosonization}

In this appendix, we present an effective theory to model the one-dimensional wire and lattice in presence of strong interactions.

We focus on the case of a single conduction channel in the wire; at zero temperature this approximation is valid as long as the Fermi energy in the wire is smaller than $\hbar \omega_\perp$ \cite{Fuchs2004}, or, equivalently,
$(\rho_0 a_\perp)^2 \lesssim 1$,
where $\rho_0$ is the 1D density including both spin species, transverse confinement frequency $\omega_\perp = \sqrt{\omega_x \omega_z} = 2\pi \cdot \SI{15.2}{\kilo\hertz}$ and $a_\perp = \sqrt{\hbar/m \omega_\perp} = 6.28 \cdot 10^3 \, a_0$ is the transversal oscillator length. In the experiment, this corresponds to densities $\rho_0 \lesssim \SI{3}{atoms/\micro\meter}$, which is fulfilled in the central region of the wire.

\begin{figure*}
    \centering
    \includegraphics[width=0.70\linewidth]{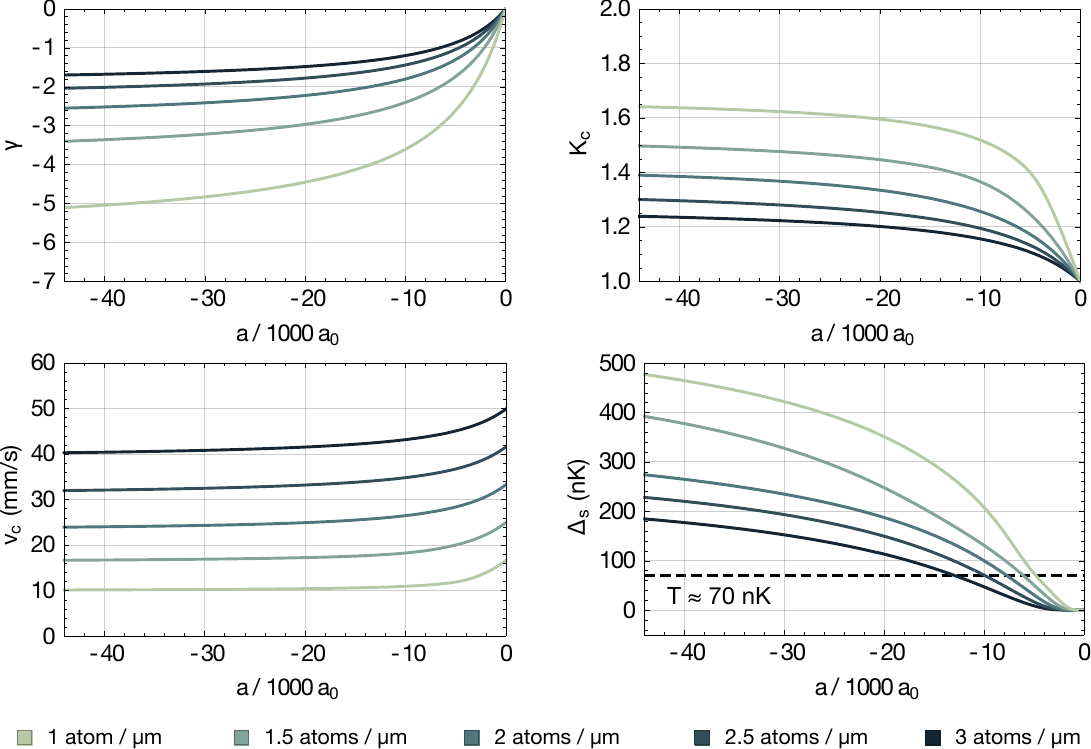}
    \caption{Gaudin-Yang parameter $\gamma$, charge sector Luttinger parameter $K_c = v_F/v_c$ , charge sector velocity $v_c$ and spin gap $\Delta_s$ as a function of 3D scattering length and linear density. The dashed line on the last panel represents the experimental temperature $T \approx 70$ nK.}
    \label{fig:vc}
\end{figure*}

In one dimension, the crossover from attractive fermions to strongly-bound repulsive bosonic dimers does not happen across the 3D unitarity point ($a=\pm\infty$), but at the confinement-induced resonance (CIR) \cite{Fuchs2004} for a positive 3D scattering length $a_\mathrm{CIR} = a_\perp / A = +6.08 \cdot 10^3 \, a_0$, with $A \approx 1.0326$. The 3D unitarity point loses its peculiarity in 1D and the 1D system is there still described by strongly attractive fermions with finite $g_1$.
In this work, all experiments are performed at $a < 0$ and are therefore on the attractive side of the CIR. Interactions are parametrized in the 1D region by an interaction strength $g_1 = 2\hbar \omega_\perp a(1-Aa/a_\perp)^{-1}$,
and a dimensionless Gaudin-Yang parameter $\gamma = mg_1/\hbar^2 \rho_0 < 0$, see Fig.~\ref{fig:vc}.

The experimentally relevant low-energy degrees of freedom of a two-species fermionic system are described by the bosonized Hamiltonian densities for the charge ($c$) and the spin ($s$) sectors \cite{giamarchi_quantum_2003}:
\begin{align}
\mathcal{H}_c &= \dfrac{1}{2\pi} \bra{v_c K_c (\nabla \theta_c)^2 + \frac {v_c} {K_c} (\nabla \phi_c)^2} - V(y) \rho(y),\label{eq:Hc}
\\
\mathcal{H}_s &= \dfrac{1}{2\pi} \bra{v_s K_s (\nabla \theta_s)^2 + \frac {v_s} {K_s} (\nabla \phi_s)^2}
\nonumber\\
&+\dfrac{2g_1}{(2\pi \alpha)^2} \cos(\sqrt 8 \phi_s)\label{eq:SG1}
\end{align}
where here and in the following we set $\hbar = k_B = 1$, $y$ is the direction along the wire, $\phi_{c,s}(y)$, $\theta_{c,s}(y)$ are the bosonic fields, $v_{c,s}$ and $K_{c,s}$ are the Luttinger parameters, $\alpha$ is the cutoff, $V(y)$ is the external potential and the density $\rho(y)$ is given by
\begin{align}
\rho(y) &= \rho_0 - \dfrac{\sqrt 2}{\pi}\nabla \phi_c(y)
\nonumber\\
&+ \rho_0\bra{
    e^{i\pa{2k_F y - \sqrt 2 \phi_c(y)}} \cos(\sqrt 2 \phi_s(y)) + c.c.
}
\nonumber\\
&+ C \rho_0 \bra{
e^{i\pa{4k_F y - 2\sqrt 2 \phi_c(y)}
} + c.c.        }.\label{eq:rho1}
\end{align}

The non-universal amplitude $C$ in \eqref{eq:rho1} is interaction-dependent and in our strongly interacting case ($|\gamma|>1$) reduces to $C\approx 1$.
The Luttinger parameters $v_{c,s}$, $K_{c,s}$ are functions of the parameter $\gamma$ and can be numerically calculated from the exactly solvable Gaudin-Yang model \eqref{eq:Gaudin-Yang} using Bethe ansatz \cite{Fuchs2004}, see Fig.~\ref{fig:vc}.
The renormalization treatment of the sine-Gordon Hamiltonian \eqref{eq:SG1} shows \cite{giamarchi_quantum_2003} that at $\gamma<0$ the spin sector flows to a strong coupling fixed point, so $\phi_s$ orders opening a spin gap $\Delta_s$, and the dispersion relation for low-lying spin excitations becomes $\e_k = \sqrt{(\Delta_s/2\hbar)^2 + (v_s k)^2}$ \cite{Fuchs2004}. Physically, this corresponds to the tendency of the fermions to form bound pairs.

Such a quantum phase with gapped spin and gapless charge modes is referred in the literature as the Luther-Emery liquid \cite{luther_backward_1974}. Like Luttinger liquids, Luther-Emery liquids have a diverging DC conductivity and can be considered as analogs of superconductors in 1D, with the formation of singlet fermionic pairs.
As long as $T<\Delta_s$, which happens already at $a\lesssim -5 \cdot 10^3 \, a_0$, the spin sector influences local densities only through Gaussian fluctuations of the $\phi_s$ field around zero, allowing us to simplify the expression \eqref{eq:rho1} to
\begin{align}
\rho(y) &= \rho_0 - \dfrac{\sqrt 2}{\pi}\nabla \phi_c(y)
\nonumber\\
&+ 2\rho_0 f_s \cos
\pa{2k_F y - \sqrt 2 \phi_c(y)}
\nonumber\\
&+ 2C\rho_0 \cos\pa{4k_F y - 2\sqrt 2 \phi_c(y)},
\label{eq:density simple}
\end{align}
where the influence of the spin sector mentioned above is taken into account by the fluctuation factor $f_s = \q{\cos\sqrt 2 \phi_s(0)} \approx 0.5$.

\section{Conductance}
\label{A:Conductance}

To calculate the conductance of the strongly-interacting fermions in a periodic potential, we follow an approach similar to the one introduced by Maslov and Stone in \cite{Maslov1995}.
The current is driven by the chemical potential difference $\Delta \mu$ between the ends of the 1D wire.
Namely, when there is an external chemical potential, the Hamiltonian density of the charge sector \eqref{eq:Hc} acquires an additional term
\begin{align}
\mathcal{H}_\mu(y) &= - \mu(y) \rho(y) = \mu(y) \dfrac{\sqrt 2}{\pi} \nabla \phi_c,
\end{align}
where we omitted the constant factors which will not show up in the equation of motion as well as the oscillating factors which will average to zero by calculating average currents.
We assume that the chemical potential takes two different, constant values in the leads, $\mu_L = \mu_R + \Delta \mu$, and changes linearly in the wire between the leads $\mu(y) = \mu_L - E y$ (the sign is chosen as to induce a left-to-right current), so that we can introduce a fictitious electric field $E = -\nabla \mu(y) = \Delta \mu/L_E$, where $L_E$ is the length of the region where the electric field is applied.

The total current
\begin{equation}
I_{\uparrow\downarrow}(y) = \dfrac{\sqrt 2}{\pi} \partial_t \phi_c(y,t)
\end{equation}
is given by the continuity equation $\partial_t \rho + \nabla j = 0$.

The one-species conductance that is relevant experimentally is
\begin{equation}
G_\uparrow = \dfrac{1}{2L} \int dy\, \dfrac{I_{\uparrow\downarrow}(y)}{\Delta \mu}.
\end{equation}

Assuming that the experimental temperature is large enough to neglect the correction due the quantum tunneling, we calculate the conductance by numerically solving the classical equation of motion for $\phi_c$, which reads:
\begin{widetext}
\begin{gather}
- 2\sqrt 2  V(y) \rho_0
\bra{2 \sin\pa{2\sqrt 2 \phi_c(y,t) - 4 k_F y} +
    f_s\sin\pa{\sqrt 2 \phi_c(y,t) - 2 k_F y}
} =\nonumber\\
=
\dfrac{v_c}{\pi K_c} \partial_{yy}\phi_c(y,t)- \dfrac{1}{\pi v_c K_c} \partial_{tt}\phi_{c}(y,t)-\dfrac{\sqrt 2 V'(y)}{ \pi}
+ \dfrac{\sqrt 2}{\pi} E(y)
.
\label{eq: of motion}
\end{gather}
\end{widetext}

We solve the equation of motion \eqref{eq: of motion} with the initial condition $\phi_c(y,0) = 0$ (no current at $t=0$) in a finite-size region of length $L = \SI{16}{\micro\meter}$. The boundary conditions are of the Sommerfeld type (outgoing-wave) with additional white noise, which takes into account the finite temperature of the leads. Namely, the boundary term is
\begin{align}
d\phi_c(L,t) + v_c\, \nabla \phi_c(L,t)\, dt &= \sigma_T\, dW_L(t),\nonumber\\
d\phi_c(0,t) - v_c\, \nabla \phi_c(0,t)\, dt &= \sigma_T\, dW_0(t),\label{eq:bc}
\end{align}
where $dW_{L,0}(t)$ are two independent Wiener processes and $\sigma_T$ is a phenomenological parameter characterizing the amplitude of the thermal noise.

The resulting stochastic differential equation is solved with the Euler-Maruyama method.
To calibrate the noise we solve the dynamics of the Luttinger liquid without external potential and without the applied voltage, i.e. we solve \eqref{eq: of motion} with $V(y)=0$, $E(y,t) = 0$, and $\phi_c(y,0) = 0$ as initial condition, but with the noisy boundary conditions \eqref{eq:bc}.
For each value of the noise amplitude $\sigma_T$ and filling factor $\nu = d \rho_0/2$, we then calculate the equal-time correlation function \cite{giamarchi_quantum_2003}
\begin{equation}
C_T(y) = \q{[\phi_c(y,t)-\phi_c(0,t)]^2},
\end{equation}
where the angular brackets denote the time average (as the Luttinger liquid with noise is ergodic and the time average is equal to the ensemble average), and zero marks the middle of the wire.

Taking this into account, the exact equal-time correlation function \cite{giamarchi_quantum_2003} in the thermal ensemble is
\begin{align}
C_T(y) &= K F_1(y) = \dfrac{K_c}{2} \log \bra{
    \dfrac{\beta^2 v_c^2}{\pi^2 \alpha^2}
    \,\sinh^2\pfrac{\pi y}{\beta v_c}
}
\nonumber\\
&\approx \dfrac{K_c}{2} \log \bra{
    \dfrac{\beta^2 v_c^2}{4\pi^2 \alpha^2}
    \,\exp\pfrac{2\pi y}{\beta v_c}}
\nonumber\\
&\approx
K \log
\pfrac{\beta v_c}{2\pi \alpha} + \dfrac{K_c\pi y}{\beta v_c},
\end{align}
where $\beta = 1/T$, and $\alpha$ is the cutoff. In the last line, we used the experimentally relevant high-temperature limit
$\pi y/\beta v_c \gg 1$.
We then fit $C_T(y)$ with a straight line $C_T(y) = A + B y$ and calculate the temperature from its slope $B = K_c \pi T/v_c$.

\begin{figure}[!htb]
    \centering
    \includegraphics[width=0.9\linewidth]{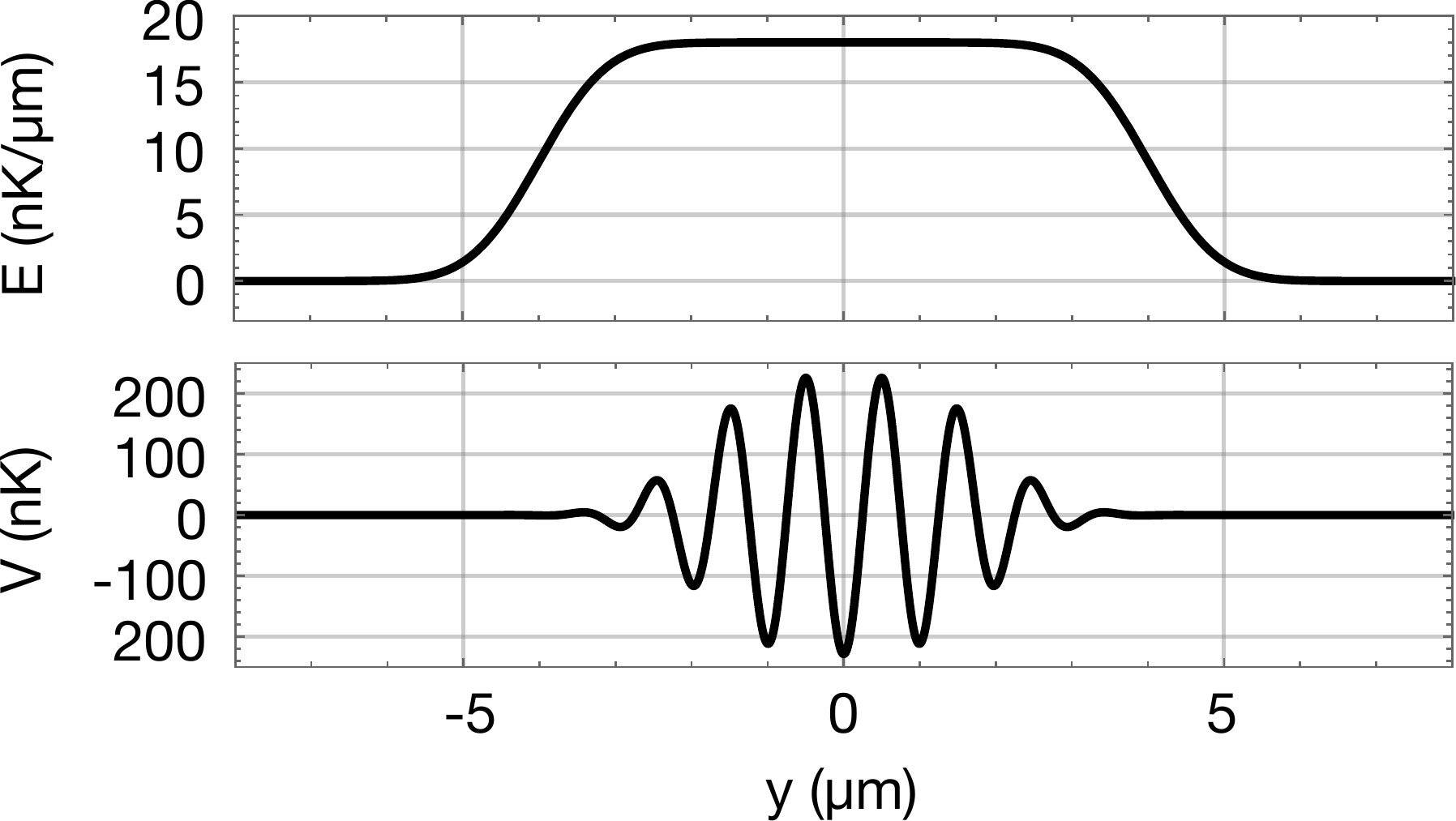}
    \caption{Profiles of the fictitious electric field (top) and the lattice potential (bottom) used in the numerical simulations.}
    \label{fig:potential}
\end{figure}

For numerical stability, the lattice and the electric field are applied in the central region of length $L_V = \SI{4}{\micro\meter}$ and $L_E = \SI{8}{\micro\meter}$ respectively; at the ends of this central region they smoothly go to zero with $\sigma_h = \SI{1}{\micro\meter}$:
\begin{align}
E(y) &= E\,\mathrm{hill}(y, L_E), \\
V(y) &= V_0[\sin(\pi y/d)^2 - 1/2]\, \mathrm{hill}(y,L_V), \\
\mathrm{hill}(y,L) &= \dfrac{1}{2}
\left[
    \mathrm{erf}\pa{\dfrac{-y+L/2}{\sigma_h}} + \mathrm{erf}\pa{\dfrac{y+L/2}{\sigma_h}}
\right],
\end{align}
where $V_0 = 459$ nK and $E = \SI{18.75}{\nano\kelvin/\micro\meter}$, see Fig.~\ref{fig:potential}.
To test the algorithm, we successfully reproduced the results of \cite{Maslov1995} in the absence of the optical lattice, with and without thermal noise.

\begin{figure*}[t]
    \centering
    \includegraphics{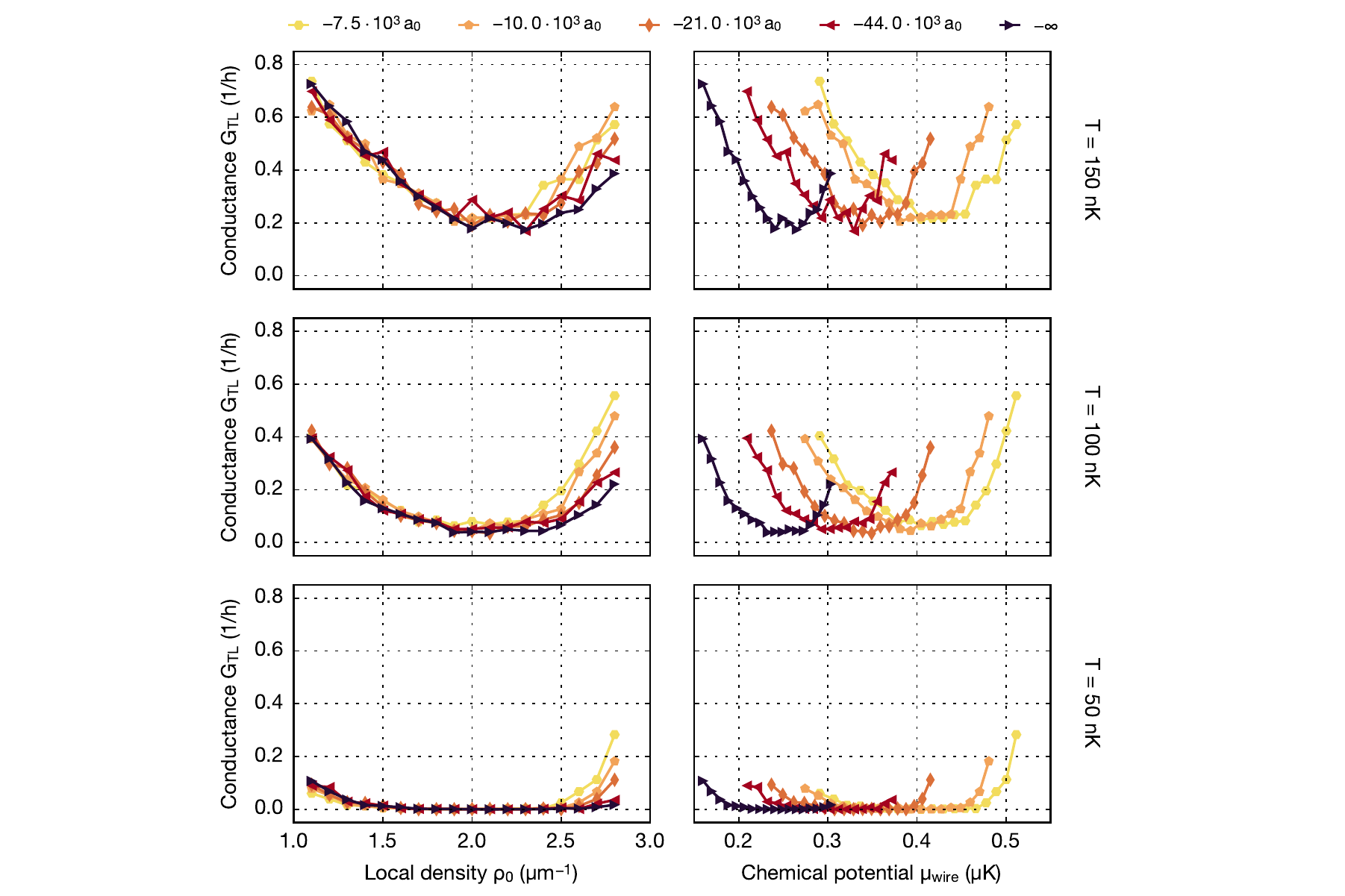}
    \caption[Results: Conductance vs $\rho_0$ and $\mu_\text{wire}$]
    {Conductance $G_\text{TL}$ obtained from Tomonaga-Luttinger theory for one spin species in units of the conductance quantum $1/h$ as a function of the mean 1D density $\rho_0$ in the wire (left) and the local chemical potential in the wire $\mu_\text{wire}$ (right) for the scattering lengths used in the experiment.
    All curves have a minimum around $\rho_0 = \SI{2}{\micro\meter^{-1}}$, which is the filling of a band insulator in the non-interacting limit.
    Stronger interactions decrease conductance and tend to `widen' the conductance minima when plotted as a function of 1D density, which can be explained by the renormalization of the potential due to spin fluctuations towards larger effective lattice heights.
    }
    \label{fig:results}
\end{figure*}

The results of the finite-temperature simulations with the optical lattice are summarized in Fig.~\ref{fig:results}, where conductance is plotted as a function of the local 1D density $\rho_0$ and the local chemical potential $\mu_\text{wire}$ in the wire. They predict a conductance minimum at a particle density $\rho_0$ of two-particles per site,  which is the density where a band insulator is formed in absence of interactions.
The local chemical potential $\mu_\text{wire}$ is obtained from $\rho_0$ using local density approximation and the zero-temperature equation of state $\mu(\rho_\text{3D}, a)$ \eqref{eq:P}.
We assume there that the 3D density at the center of the atom cloud is given as $\rho_\text{3D} = \rho_0 / \sigma_\text{2D}$, where $\sigma_\text{2D} \approx \SI{1}{\micro\meter^2}$ is the estimated wire cross-section.
Although it does not account for the complex geometry of the constriction, which continuously transforms 3D reservoirs into 2D sheets and then to the 1D wire, this simplified model is able to produce interaction-dependent shifts of the conductance minimum along the chemical potential axis that are quantitatively comparable to the experimental data.

Our Luttinger liquid formalism is only applicable when a single conductance channel contributes to transport, which happens in practice for chemical potentials close to the conductance minimum.
At low chemical potentials $\mu_\text{wire} < 0.2 - \SI{0.3}{\micro\kelvin}$, it does not capture the drop of conductance associated with the closure of a single conductance channel \cite{husmann_connecting_2015}. 
On the other hand, for larger chemical potentials $\mu_\text{wire} > 0.4 - \SI{0.5}{\micro\kelvin}$, the high conductance experimentally observed might be explained by the contribution of the higher transverse channels due to Andreev processes, as put forward in \cite{Kanasz-Nagy2016} to account for anomalous conductances observed in a quantum point contact with attractive interactions \cite{krinner_mapping_2016}.

The chemical potential $\mu_\text{wire}$ can be related to the experimental variable $V_g$ through
\begin{equation}
V_g + \mu_\text{res} =  V_\text{wire} + \mu_\text{wire},
\label{eq:mu_res}
\end{equation}
where $V_\text{wire}$ is the local ground-state energy at the wire center, and $\mu_\text{res}$ is the chemical potential away from the constriction inside the reservoirs.
If no spurious light potential is present in the wire, the ground-state energy $V_\text{wire}$ is about the zero-point energy of the wire transverse confinement, $V_\text{wire} = \hbar(\omega_x + \omega_z)/2 = \SI{830}{\nano\kelvin}$, up to a renormalization factor due to interactions. By equating $V_g$ to the gate potentials where the conductance minima are experimentally observed (cf. Fig.~5) and $\mu_\text{wire}$ to the theoretical values corresponding to a commensurate filling of two atoms per site, the reservoir chemical potential can be estimated from \eqref{eq:mu_res} to be $\mu_\text{res} \sim 230 - \SI{260}{\nano\kelvin}$.

\section{Luther-Emery liquid and super-Tonks-Girardeau gas}
\label{A:Super-Tonks-Girardeau gas}

The Cooper pairs become effectively unbreakable molecules only at the CIR, where the system can be described by a Tonks-Girardeau gas of bosons. In our experiment we explore the regime of moderately strong attractions between the fermions, where the pairs have a finite size and are tightly bound, but not unbreakable.

In this section we show that in the absence of a periodic potential a fermionic gas in Luther-Emery phase can be mapped to a so-called super-Tonks-Girardeau (STG) gas, a gas of bosons featuring long-range repulsion and a Luttinger parameter $K_b<1$ (the subscript `$b$' stands for `bosonic').

First we note that as long as the spin sector is ordered (gapped) at $\q{\phi_s} = 0$, we need to deal with the Luttinger liquid of spinless fermions in the charge sector only, described by the bosonized fields $\phi_c(y), \theta_c(y)$ to obtain correctly the low energy physics of the full model.

The bosonized density operator of 1D spinless bosons is given by \cite{cazalilla_one_2011}
\begin{equation}
\rho_b(y) = \pa{\bar \rho_{b} - \dfrac{1}{\pi} \nabla \phi_b(y)}\sum_{\ell = -\infty}^{\infty} e^{2i\ell \bra{\pi \bar \rho_{b} y - \phi_b(y)}}, \label{eq: density bosons}
\end{equation}
where $\bar \rho_{b}$ is the average bosonic density.

\begin{figure}[!htb]
    \centering
    \includegraphics{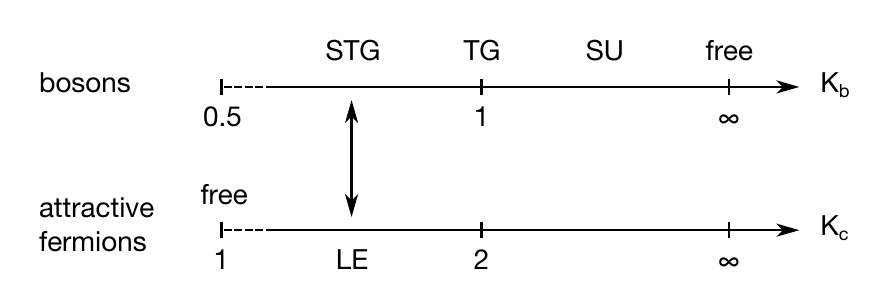}
    \caption{Phase diagram of 1D bosons (top) and 1D spinless attractive fermions (bottom) with respect to the Luttinger parameter $K_{b,c}$. The vertical arrow represents the mapping of the attractive fermions in the Luther-Emery (LE) phase to the Super-Tonks-Girardeau gas (STG) discussed in the text. Bosons realize the STG phase at $K_b < 1$, hard-core Tonks-Girardeau (TG) phase at $K_b = 1$ and a superfluid of weakly repulsive bosons (SU) at $K_b \gg 1$; the free boson limit is realized at $K_b \rightarrow \infty$. The fermions are repulsive for $K_c<1$, free at $K_c=1$ and attractive at $K_c>1$. The dashed region of the phase diagram corresponds to the regime where the boson-to-fermion mapping breaks down due to fluctuations of the spin field.}
    \label{fig:phase diagram}
\end{figure}

Then comparing \eqref{eq: density bosons}
with the one for 1D attractive fermions \eqref{eq:density simple} and taking into account only the relevant harmonics $\ell = \{0, \pm1, \pm2\}$, we can establish an approximate correspondence of our attractive fermions and interacting bosons:
\begin{align}
\rho_0 & \leftrightarrow 2 \bar \rho_{b}, \nonumber\\
\phi_c & \leftrightarrow \sqrt{2} \phi_b, \nonumber\\
k_F & \leftrightarrow \pi \bar \rho_{b}, \label{eq:mapping}
\end{align}
from which it follows that $K_c \leftrightarrow 2K_b$ and enables us to draw a concurrent phase diagram of the 1D bosons and fermions, see Fig.~\ref{fig:phase diagram}.

Experimentally, attractive interactions between fermions stronger than $|a| > 10^4 \,  a_0$ are routinely achieved, which corresponds to the fermionic Luttinger parameter $K_c \approx 1.4$, see Fig.~\ref{fig:vc}. This allows us to map our system to a STG bosonic gas of $K_b \approx 0.7$. Through this mapping, all long-wavelength correlation functions for the Cooper pairs can be directly mapped to the bosonic ones for the STG gas.

This correspondence holds only as long as
the spin sector is well locked in its minimum (implying that $f_s = \q{\cos\sqrt 2 \phi_s(0)} \lesssim 1$) so that the Fermi gas is described by one density mode. For instance, this is fulfilled in the Tonks-Girardeau gas, where $f_s = 1$ and our mapping \eqref{eq:mapping} of Cooper pairs to bosons is exact. If the interaction becomes too weak, the spin gap becomes lower than the energy scale at which the system is probed, and then it is necessary to include the two modes (spin and charge). In this regime one cannot map the system to a single-mode bosonic STG gas anymore.

\begin{figure*}[t]
    \centering
    \includegraphics[width=0.7\linewidth]{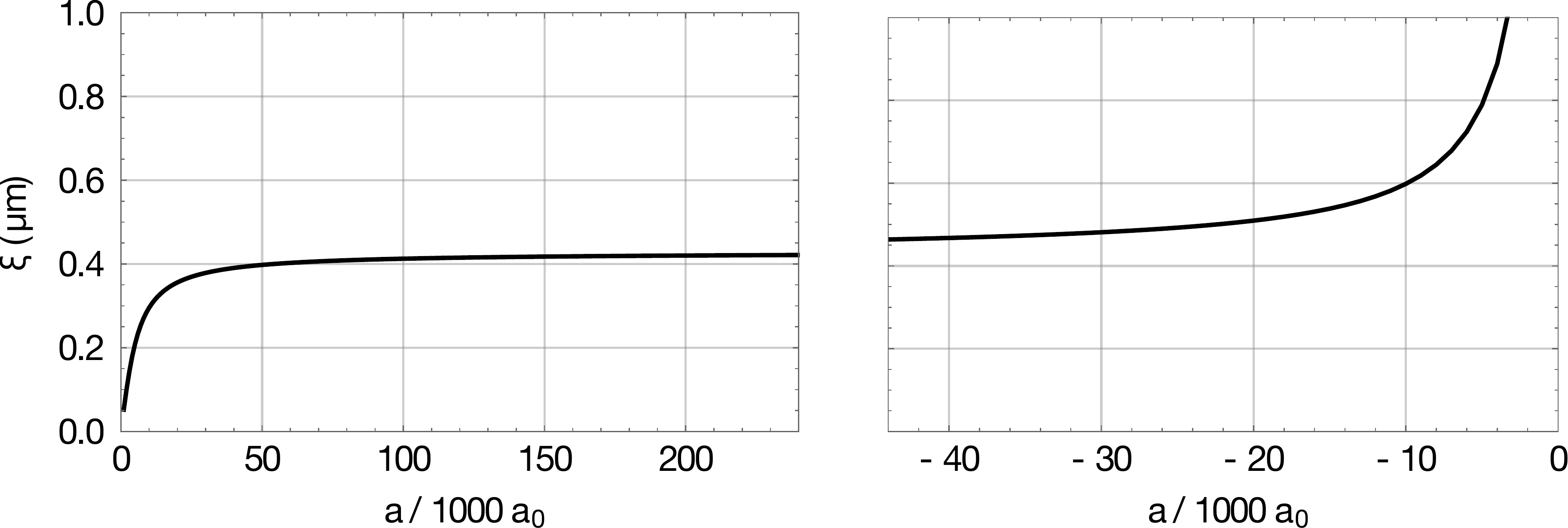}
    \caption{Pair size $\xi$ as a function of the scattering length. Linear densities $\rho_0$ from 1 to $\SI{3}{atoms/\micro\meter}$ correspond to interparticle spacings from 1 to about $\SI{0.3}{\micro\meter}$, which is of the order of the pair size. For higher densities and/or weaker interactions the pairs become larger than the interparticle separation and the mapping to a Super-Tonks-Girardeau gas fails. Left and right panels correspond to positive and negative scattering lengths $a$.}
    \label{fig:xi}
\end{figure*}

We estimate the pair size as $\xi = \hbar/\sqrt{m \e_0}$, where $\e_0$ is the energy of the bound state \cite{tokatly_dilute_2004}, see Fig.~\ref{fig:xi}. At moderately strong attraction and densities $\rho \approx \SI{2}{atoms/\micro\meter}$, the pair size $\xi \approx \SI{0.4}{\micro\meter}$ is comparable but smaller than the interparticle separation $\rho^{-1} \approx \SI{0.5}{\micro\meter}$, and it becomes meaningful to interpret the pairs as localized bosons. However, we note that the mapping to a STG gas continues to hold when the pair size is larger than the inter-particle spacing, as it only relies on the locking of the spin field. In the latter case, the excitations are density waves and cannot be interpreted as tightly bound pairs.

\bibliographystyle{apsrev4-1}
\bibliography{paper}

\end{document}